\documentclass{aa}
\usepackage{graphicx,psfig}
\usepackage{aalongtable,lscape}

\newcommand{\Msun} {M$_\odot$}
\newcommand{\Lsun} {L$_\odot$}

\newcommand{\Tstar} {T$_{\rm{eff}}$}
\newcommand{\Lstar} {L$_\star$}
\newcommand{\Mstar} {M$_\star$}

\newcommand{\mr} {M$_\star$/R$_\star$}

\newcommand{\AJ} {A$_J$}
\newcommand{\um} {$\mu$m}

\newcommand{\Myr} {M$_\odot$/yr}

\newcommand{\Ha} {H$\alpha$}

\newcommand{\Roph} {$\rho$~Oph}
\newcommand{\Pab} {Pa$\beta$}
\newcommand{\Brg} {Br$\gamma$}

\newcommand{\Macc} {$\dot M_{acc}$}
\newcommand{\Lacc} {L$_{acc}$}

\newcommand{\simless}{\mathbin{\lower 3pt\hbox
      {$\rlap{\raise 5pt\hbox{$\char'074$}}\mathchar"7218$}}} %< or of order
\newcommand{\simgreat}{\mathbin{\lower 3pt\hbox
     {$\rlap{\raise 5pt\hbox{$\char'076$}}\mathchar"7218$}}} %> or of order

\begin{document}

\title{Accretion  in the $\rho$-Oph pre-main sequence stars
\thanks{
Based on observations collected at the European Southern Observatory, Chile.  Program 073.C-0179.}
}

\author{
A. Natta\inst{1},
L. Testi\inst{1} 
\and
S. Randich \inst{1}
}
\institute{
    Osservatorio Astrofisico di Arcetri, INAF, Largo E.Fermi 5,
    I-50125 Firenze, Italy }

\offprints{natta@arcetri.astro.it}
\date{Received ...; accepted ...}

\authorrunning{Natta et al.}
\titlerunning{Accretion in $\rho$-Oph}

\abstract 
{}
{The aim of this paper is to provide a measurement of the mass accretion rate
in a large, complete sample of objects in the core of the
star forming region \Roph.} 
{The sample includes most of  the objects (104 out of 111)
with evidence of a circumstellar disk
from mid-infrared photometry; it covers a stellar mass range
from about 0.03 to 3 \Msun\ and it is  complete to a limiting mass
of $\sim$0.05 \Msun.  
We used J and K-band  spectra to derive the mass accretion rate of each object from the
intensity of the hydrogen recombination lines, \Pab\ or \Brg.
For comparison, we also obtained similar spectra of
35 diskless objects.}
{The results show that emission in these lines is only seen in stars
with disks, and
can be used as an  indicator of accretion. 
However, the converse does not hold, as about 50\% of our disk objects do not
have detectable line emission.
The measured accretion rates show a strong correlation with the mass of the central object (\Macc$\propto M_\star^{1.8\pm0.2}$)
and a large spread, of two orders of magnitude at least, for any  interval of \Mstar. A comparison with existing data for Taurus shows that the 
objects in the two regions have  similar behaviour, at least for objects
more massive than $\sim 0.1$ \Msun. The implications of these results are
briefly discussed.}
{} 

\maketitle
\section {Introduction}
Accretion disks are common around young stars of all mass,
from \Mstar$\sim$3 \Msun\ down to
very low mass objects and brown dwarfs.
They form during the collapse of the molecular core, from which the star
is born, and last well beyond this initial phase, when the core has dispersed and the star has acquired most of its final mass. 
The accretion disks are the birthplace of planets,
whose formation and evolution are controlled by the disk physics.

Even if  accretion disks have been part of the accepted paradigm of star
formation 
for many years, many of their physical properties are 
poorly known, and  the physical
mechanism of angular momentum transfer, which determines the disk
evolution, is still unclear.
%Telescopes of the ten-meter class are now available to
%study of the properties and evolution of disks for a large
%sample of objects over a broad interval of masses
%in different regions
%of star formation, as
%required  to investigate
%the disk physics and its relations to planet formation.
The physical quantity
that controls the accretion phase is the mass accretion rate
through the disk \Macc. This quantity can be derived only indirectly, by
fitting models to observed quantities such as the UV excess emission
and/or the profiles and intensity of lines
believed to form in the accreting gas. Measurements of \Macc\
are now available
for a large number of stars in Taurus
(e.g., Muzerolle et al.~\cite{Mea05} and references therein).
The results  have shown that \Macc\ is a
strong function of the mass of the central object, roughly
$\propto M_\star^2$, and that  a
large dispersion is present (about two orders of magnitude)
for objects with the same \Mstar. 
%This does not include weak-line or diskless TTS, which have
%no evidence of accretion-related activity whatsoever.
%The dependence of \Macc\ on \Mstar\ and its large dispersion
%for objects of similar  age and mass 
Both results are  a challenge for
accretion disk models, as discussed, e.g., by Muzerolle et al.~(\cite{Mea03})
and Natta et al.~(\cite{Nea04}).

Measurements of  accretion rates in other star forming regions
are scarce in comparison, mostly limited to very low mass objects
(Muzerolle et al.~\cite{Mea03}, \cite{Mea05}).
In a study of very low mass objects and brown dwarfs in
Ophiuchus, 
Natta et al.~(\cite{Nea04})
found that  they are actively accreting
with \Macc\ higher by at least one order of magnitude
than  objects of similar mass in Taurus. This could be due
to a difference in age, since the Ophiuchus BDs
are  very young objects, younger than their Taurus counterparts,
but could also be due to different environmental
conditions.   
%To improve our understanding of the
%accretion process, Natta et al.~(\cite{Nea04}) pointed out the
%importance to determine accretion rates of objects 
%covering a large range of stellar masses
%in different star-forming regions.

While it is clearly necessary to improve the physical
models of accretion disks,
at the same time it is important to
study large and if possible complete samples of stars
in a variety of star forming regions,  differing
in age and  global properties.

We report in this paper the results of a project
aimed at measuring
the mass accretion rate of a large sample of pre-main
sequence objects, ranging  from a few solar masses
to few tens of Jupiter masses, 
in the star forming  region \Roph.
The core of
\Roph\ is perfectly suited for such a study, as
it is rich in pre-main sequence stars,
which include intermediate mass objects, T Tauri stars (TTS)
and brown dwarfs (BDs).  Its
stellar content has been studied, e.g., by
Luhman \& Rieke (\cite{LR99}; LR99 in the following,  and references
therein to previous work), Natta et al.~(\cite{Nea02})
 and, more recently,
by Wilking et al.~(\cite{Wea05}).
 Moreover, \Roph\  is  very different from Taurus,
 younger
and more compact, and it  will allow us to explore    
 the accretion properties of pre-main sequence stars under different conditions,
following the results of Natta et al.~(\cite{Nea04}).

Ophiuchus
 has been observed in two mid-IR bands with ISO by
Bontemps et al.~(\cite{BKA01}; BKA01 in the following),
who detected 199 sources in the \Roph\ core.
Of these, 111 were classified, on the basis of their IR colors, as Class II
objects, i.e.,  visible young stellar objects with evidence of disks.
They provide a sample of systems with disks {\it complete}
to a limiting mass of about 0.05 \Msun.
In a  spectroscopic study of the very low luminosity
objects of the BKA01 sample, 
Natta et al.~(\cite{Nea02})
confirmed that they were BDs with mid-IR excess, very likely from
a circumstellar disk; as mentioned, these BDs
show significant differences in  accretion properties
from their analogs in Taurus. 

%It is therefore important to study other
%star forming regions, in addition to Taurus, to check how
%general these results are, how accretion vary with time and environment,
%to which degree the disk properties and evolution  are controlled
%by the properties of the pre-stellar cores from which they form.
%Statistically, there is evidence
%that  accretion-related
%activity must fade with age, as the fraction of stars with
%circumstellar disks decreases 
%(e.g., Haisch et al.~\cite{Haea01}, \cite{Haea05} and references therein).
%However, it is difficult to identify any  trend
%between time, as measured by the stellar age, and the activity
%level for objects within the same star forming region.
%A classical example of this is the well know result that in Taurus
%Classical (i.e., disk accretors) TTS  and weak-line (i.e.,
%diskless and non-accreting) TTS have similar ages (ref?).

%Our aim was to confirme the Taurus results and to
%see, more precisely, if the  range of \Macc\ and its dependence on
%\Mstar\ vary with the age of the region, or with its physical
%caracteristics, such as its density.

The disadvantage of observing \Roph\ is its high extinction, which makes
veiling measurements in the UV and visual impossible except for a few objects.
The most effective way to determine \Macc\ for the Ophiuchus sample
is therefore
to use  the luminosity of
hydrogen recombination lines, such as \Pab\ and/or \Brg.
The relation between IR line luminosity and  accretion luminosity,
independently measured from the UV excess, was established by
Muzerolle et al.~(\cite{Mea98b}) for TTS, and by Calvet et al.~(\cite{Cea04})
for intermediate mass objects.  Natta et al.~(\cite{Nea04})
 extended it to very low mass objects, where \Macc\ was determined by
fitting the observed \Ha\ profiles with the predictions of magnetospheric
accretion models.

In this paper, we present the results of a spectroscopic
IR survey of Ophiuchus objects.
In Sec.~2, we  describe the properties of the observed sample,
which includes almost all (104 out of 111) 
the Class II  objects and
a subset (35 objects out of 77) of the diskless systems
(Class III), also from the  BKA01 survey, that we will use for comparison.
The observations, data reduction and method of analysis are
discussed in Sec.~3. The results are presented in Sec.~4
and  discussed in Sec.~5.  Sec.~6 summarizes our conclusions.

\section {Characteristics of the observed  sample}

\subsection {The sample}

The most complete survey of young stellar objects in the \Roph\ Main
Cloud (L~1688) is that obtained in two mid-IR bands (6.7 and 15.3 \um)
with ISOCAM (BAK01). Based on the near and mid-IR colors, the
objects were divided in Class I (accreting protostars), Class II
and tentative Class II 
(pre-main sequence stars with IR excess typical of disks, like
classical T Tauri stars or CTTS), and Class III/tentative Class III (objects with colors
typical of stellar photospheres, like weak-line T Tauri stars or WTTS).
BAK01 estimate that their Class II
sample of 111 objects
is complete to a limiting luminosity \Lstar$\sim$0.03 \Lsun,
corresponding 
approximately to 0.05 \Msun. The Class III sample is
only complete to $\sim$0.2 \Lsun\ (about 0.15 \Msun). 
Note  that not all the Class III objects have been
confirmed as \Roph\ members. 
Barsony et al.~(\cite{BRM05}) have recently confirmed the
accuracy of the ISOCAM results with  ground-based  10$\mu$m
observations of a large subset of the BKA01 sources.

Our sample includes  104 of the 111  Class II/tentative Class II objects
(Class II for simplicity in the following) listed by BAK01 in the \Roph\ core.
Most of the spectra (96) were obtained  in the J band; the remaining
8, of objects too weak in J, in the K band; one object has been
observed at both wavelengths.
As a comparison sample, we observed 35
of the 77 Class III and tentative Class III
(in the following, Class III)
objects, 31 in the J band and 4 in K.
The objects and their properties are listed in 
%Tables~\ref{table_classII} and ~\ref{table_classIII} 
Tables C.1 and C.2 \footnote{All the tables are available in electronic form only.}.

\subsection {Stellar parameters}

The stellar properties (i.e., spectral type, luminosity, mass and radius)  
of the BAK01 sample are well  known  only for a handful of objects.
The main difficulty comes from the large uncertainties that affect
spectral types, due to the combination of high extinction and
large veiling, even at near-IR wavelengths
(e.g., LR99
and references therein; Doppman et al.~\cite{Dea03}; Wilking et al.~\cite{Wea05}).
LR99,
using K-band low resolution spectra,  provide  spectral types for 
37  of our Class II objects. However, 23 of them have uncertainties of 
almost one spectral class.

Given the uncertainties, and considering that 
most of our objects do not have
any spectral classification, 
we have decided to adopt a statistical approach, following
BAK01.

First, we compute the extinction toward each object from the
observed  (J-H)-(H-K) colors, as given by 2MASS
\footnote{This publication makes use of data products from the Two Micron All Sky 
Survey, which is a joint project of the University of Massachusetts and 
the Infrared Processing and Analysis Center/California Institute of 
Technology, funded by the National Aeronautics and Space Administration 
and the National Science Foundation.
}, corrected to CIT system,
adopting the Ophiuchus extinction  law of Kenyon et al.~(\cite{KLB98})
and the locus of CTTS defined by Meyer et al.~(\cite{MCH97}).
The result can be expressed as:

\begin{equation}
A_J = 2.31 \,\big( 1.72\,(J-H)_{CIT} -(H-K)_{CIT} -0.896\big)
\end{equation}

\noindent This relation gives the correct reddening also for objects
with no excess in the near infared, as long as they have \Tstar $\simless
5000$ K (spectral type later than K2), which applies to
the majority of our sample.
However, for diskless objects of earlier spectral type
it will underestimate \AJ\ and thus the inferred
\Lstar\ significantly. We will come back to this point in \S 4.1.

The stellar luminosity is  computed from the J magnitude and \AJ,
using a bolometric correction  similar to that adopted by BAK01:
%slightly modified to improve the agreement between our
%estimates and those of LR99:

\begin{equation}
\log{L_\star} \> =\> 1.24 \> + \> 1.1 \log{L_J}
\end{equation}
where  \Lstar, $L_J$ are in units if \Lsun\ and $L_J = 301 \times 10^{-(J-A_J)/2.5}$. 
 Eq.(2) assumes that the J-band  disk emission is negligible in comparison to the photospheric one. If this is not the case, the equation
overestimates \Lstar. A recent work by Cieza et al.~(\cite{Cea05})
suggests that classical TTS have a J-band excess of $\sim 0.3$
mag on average, so that  \Lstar\ derived from the J magnitude
is  higher than the true one by about 30\%.
We have checked that, if such a correction applies
to objects of all masses,
none of our conclusions will change.
Our values of
\Lstar\ are in agreement with LR99 estimates  always within a factor of 2, and generally much better.
We have also compared the luminosities derived in this way with the
results obtained by Natta et al.~(\cite{Nea02}) from near-infrared
J,H,K low resolution spectroscopy for a group of 10 BDs;
also in this case, the results are within a factor of 2, with the exception of
one object ($\rho$Oph-ISO 033), for which   we underestimate the luminosity by almost
one order of magnitude. 
This discrepancy has
no impact on the results of this paper.

When only H and K magnitudes were available (17 Class II and 1 Class III
objects), we  estimated the stellar luminosity
using eq.(2) and (4) of BKA01.

There are 5 objects (4 Class II and 1 Class III)
that have companions clearly seen in our spectra, but
which are not resolved in the 2MASS
photometry. All  the  companions
have a good detection of the continuum; the
flux ratio between the primary and the secondary 
is always larger than a factor of 3.
Two of the companions ($\rho$Oph-ISO 068b and $\rho$Oph-ISO 072b)
have been detected in the K-band \Roph\ multiplicity survey
of Ratzka et al.~(\cite{Ratzka05}), with flux ratios to the
primary of 0.19 and 0.16, respectively. We have
accordingly
not corrected the 2MASS magnitudes of the
primaries for the contribution of the companions, because
the corrections to the derived parameters 
would have  been within the uncertainties.
The secondary components have no detectable \Pab\ emission, and we will
omit them from our analysis in the following;  their properties
are summarized in Table C.3.
%~\ref{table_companions}.

To determine stellar radii and masses, we make the  assumption
that the star formation in Ophiuchus is coeval, and that all the objects
lie  on a single isochrone in the HR diagram. 
With this assumption, we can derive stellar mass, temperature and radius
from the measured \Lstar.
This procedure is  reasonable  for the Ophiuchus core,
whose age estimates range
between 0.5 and 1 Myr, with very few
stars older than that (BAK01; LR99). 
In the following, we adopt the D'Antona
and Mazzitelli (\cite{DM97} and 1998 web updates; DM98 in the following ) evolutionary tracks for an age of 0.5My.
The uncertainties introduced by the assumption of coeval star formation
and the differences expected if other   evolutionary tracks were
used are 
discussed in  Appendix A.

The  values of the stellar parameters are given
in Tables C.1 and C.2.

\section {Observations and data analysis}

\subsection {Observations and data reduction}

Near infrared moderate resolution J and K band spectroscopic observations
of all targets in our sample were obtained at the ESO Observatories in 
Chile. The objects were either observed using the SofI instrument at the 
NTT 3.6m telescope (June 2004, Visitor Mode) or the ISAAC instrument
at the Antu 8.2m VLT unit telescope (Spring 2004, Service Mode),
as specified in Tables C.1  and C.2.
%Tables~\ref{table_classII} and \ref{table_classIII}.
Detailed descriptions of both these instruments are available on the 
ESO web pages
\footnote{http://www.ls.eso.org/lasilla/sciops/ntt/sofi/\nolinebreak
and
http://www.eso.org/instruments/isaac/}. For all the objects that were
observable at J-band, with SofI we used the 0.6 arcsecond slit
and the Blue low resolution grims, resulting in a spectral resolution of 
approximately $\lambda/\Delta\lambda\sim 1000$ and a spectral coverage from 
$\sim 0.95$ to $\sim 1.64~\mu$m; with ISAAC we employed the short-wavelength
low resolution spectral mode with central wavelength $1.25~\mu$m and 
$0.6\arcsec$ slit width, giving a spectral resolution of
$\lambda/\Delta\lambda\sim 900$ and a spectral coverage limited to the J-band.
A number of objects were only observable at K-band, for these we either used 
the SofI Red low resolution grism with similar spectral resolution as for the 
Blue grism observations and spectral coverage from $\sim 1.6$ to $2.5~\mu$m,
or the ISAAC short wavelength low resolution mode with central wavelength
$2.2~\mu$m, which offers a similar spectral resolution as the J-band
observations and a spectral coverage limited to the K-band. Integration
times varied from about 0.5 to 2 hours on source,
depending on the expected brightness of the objects
and observing conditions (in Visitor Mode).

During the Visitor Mode observations at the NTT telescope, we acquired
several telluric standard stars per night at varying airmasses; each 
Observing Block from our programme executed in Service Mode at the VLT
was preceded or followed by a telluric standard observed with the same
instrument mode and at a similar airmass as our target stars. Spectroscopic
flat fields and arcs were obtained during daytime either before or after
our observations. Standard methods were employed to calibrate our data.
We did not attempt to obtain flux calibrated spectra; all our spectra
are wavelength calibrated using OH airglow lines and corrected on 
an arbitrary intensity scale for telluric
absorption and instrument response using the telluric standard star observations.

Correction for telluric absorption and instrumental response was 
obtained observing at similar airmasses
early type stars (early B or O) of known spectral 
type from the telluric standards lists of ISAAC 
\footnote{http://www.eso.org/instruments/isaac/tools/spectroscopic
\_standards.html}. 
These stars all have \Pab\ or \Brg\ absorption which were 
manually removed from the spectra before applying the correction.

Most of the spectra are of excellent quality; the detection limits of the
Pa$\beta$ or Br$\gamma$ equivalent width are in general of 
yhe order of 0.5--1\AA. Variations 
around this limit are mainly related to the signal to noise ratio achieved on
the photospheric continuum of the individual objects.
The signal to noise ratio depends on the telescope/instrument used, 
the observing conditions, the 
integration time and the apparent magnitude of the object.
It is not necessarily a function
of the object intrinsic luminosity because the extinction  can be very different
and because we tried as much as possible to observe 
two objects at the same time
by properly aligning the slit, so that some relatively bright source 
near a faint one
may have been observed with ISAAC and a long integration time. 
However,  most of the lower luminosity objects have been 
observed with ISAAC 
and, expecting lower line intensities, with a higher signal to noise ratio;
thus, the line detection limits for low luminosity objects are
generally lower than for intermediate luminosity ones.

The sample studied in this paper includes also the 9 BKA01 sources
for which Natta et al.~(\cite{Nea04}) obtained J and K band
spectra with ISAAC. We have 
taken the Natta et al.~(\cite{Nea04}) J band spectra and
reanalyzed them in the same manner used for the others.

%{\bf Observed profiles in Appendix A}.

\subsection {Method}

The luminosity of \Pab\ and \Brg\ are computed from the measured
equivalent widths of the emission lines
and the broad-band J and K fluxes, corrected
for extinction, determined as described in \S 2.2.
No correction for underlying photospheric absorption was applied,
since the expected equivalent width is small ($\simless 0.5$ \AA; Wallace et al.~\cite{Wea00})
for objects with \Tstar $\simless 5000$ K,
which represent the quasi-totality of our sample (see 
%Table~\ref{table_classII}) 
Table C.1 and would not change the results.

There are 12 Class II (11 of them have no
\Pab\ detection)  for which it was not possible to determine
line fluxes, due to lack of J  magnitudes; 
they  will not be
included in the following discussion. Similarly, we will not
consider further the one Class II ($\rho$Oph-ISO 035) with
weak \Pab\ in absorption. 

The accretion luminosity of each Class II object is 
deived from 
the empirical correlation between \Lacc\ and the luminosity of \Pab\ or
\Brg, derived by Natta et al.~(\cite{Nea04})
 and Calvet et al.~(\cite{Cea04}), respectively (see also Muzerolle et al.~\cite{Mea98b}):

\begin{equation}
\log{L_{acc}/L_\odot} = 1.36\, \log{L(P_\beta)/L_\odot} + 4
\end {equation}

\begin{equation}
\log{L_{acc}/L_\odot} = 0.9\, (\log{L(Br_\gamma)/L_\odot}+4)\> -0.7
\end{equation}

These relations have been calibrated using  accretion
luminosities derived by fitting the measured veiling 
(for T Tauri stars) and/or H$_\alpha$  profiles
with the predictions of magnetospheric accretion models;
the  objects used for the calibration cover
the  mass  range from $\sim 3$ \Msun\ to  brown dwarfs.
The mass accretion rate is then computed from \Lacc\ 
($\dot M_{acc}=L_{acc} \, R_\star/(GM_\star)$).
The results are given in Table C.1.
%Table~\ref{table_classII}. 

The reliabilty of our procedure  was 
verified by applying it to a sample of well studied pre-main
sequence stars, covering roughly the same range of masses, for
which reliable values of the stellar parameters (i.e.,
mass and radius) and of the accretion rate
could be found in the literature. Using literature measurements
of the \Pab\ intensity and of \Lstar,
we  derived for each object  mass and accretion rate
as  done for the Ophiuchus objects, and  
compared them to the ``real" values. Details
can be found in the Appendix B.

We have applied a similar procedure to the Class III objects; the
results are shown in Table C.2.
%Table~\ref{table_classIII}.

%Measurements or upper limits for \Macc\ sre givem in Table~1, Column ??,
%for a total of ?? Class II and ?? Class III objects.
%Missing J/K photometry: very few objects with detected lines.

\section {Results}

\subsection {Equivalent Widths}

\begin{figure}[ht!]
\begin{center}
\leavevmode
\centerline{ \psfig{file=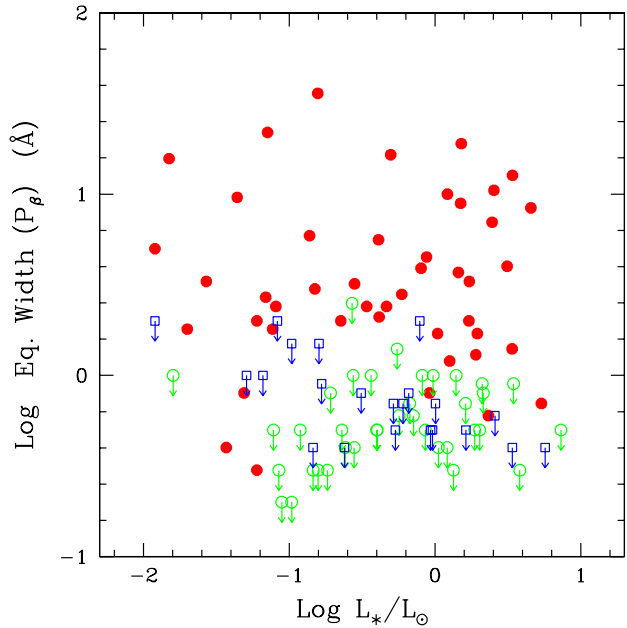,width=9cm,angle=0} }
%\centerline{ \psfig{file=ps.eqw_pab_limits,width=9cm,angle=0} }
\end{center}
\caption{Equivalent width of the \Pab\ emission line for all objects
with known \Lstar. Circles are Class II sources,
filled circles are detections,
empty circles with arrows upper limits. Squares are Class III sources.
Two objects with \Lstar$< 0.01$ \Lsun\
and  \Pab\ undetected are not plotted.}
\label{pab_width}
\end{figure}

Emission in the near-IR hydrogen recombination lines has been
detected in 45\% of Class II sources,
46 of the 96 observed in \Pab\ and 1 out of 9 observed in
\Brg.
In contrast, no Class III source shows emission in these hydrogen lines;
8 Class III objects have \Pab\ or \Brg\ in absorption and for the others we 
do not see the lines.  
The measured equivalent widths are given in 
Table C.1 and C.2.
%JTable~\ref{table_classII} and \ref{table_classIII}.
Fig.~\ref{pab_width} shows the \Pab\ equivalent width
as function of \Lstar. 

Six Class III objects have \Pab\ in absorption with equivalent widths
$\simgreat$1 \AA, i.e., larger than one can expect in late-type stars
(Wallace et al.~\cite{Wea00}). They are likely  earlier type stars, and
this is certainly the case  of \Roph-ISO 180,
which is classified A7 by Wilking et al.~(\cite{Wea05})  and of \Roph-ISO 113,
earlier than F8 according to LR99.
For these six stars, as already mentioned, the method used to estimate
\AJ\ and all the derived stellar parameters is  not correct; therefore, we
omit their stellar parameters from  Table C.2. 
%Table~\ref{table_classIII}. }

The comparison between the Class II and Class III samples
clearly  shows that
emission in the near-IR hydrogen lines, in contrast
to that in optical lines such as \Ha\, is  restricted to objects with circumstellar disks, and
can therefore be  used as a reliable accretion indicator.
However, one should keep in mind that the opposite is not
necessarily true, as about 50\% objects with disks have no detected
emission.  
%For example, our spectroscopic survey ``looses" about 1/2 of the disks.
%One star ($rho$Oph-ISO 035 or GY15) has weak \Pab\ absorption;
%it is a new Class II object in BKA, and one may wonder if its
%classification is correct.

%Inspection of Fig.~\ref{pab_width} shows 
%that there is no correlation between the equivalent width of the
%emission lines   and \Lstar.
%although
%objects  with  \Lstar$\simgreat 1$ \Lsun  have, on average,
% higher values of EW(\Pab).
The fraction of Class II objects with detected \Pab\  emission
varies from 56\% for \Lstar$\simgreat 1$ \Lsun to
42\% for $0.03\simless L_\star\simless 1$ \Lsun.
Very low luminosity  objects (7 objects with \Lstar $\simless 0.03$ \Lsun)
have a marginally higher  detection rate
($\sim$57\%), 
due in  part to the sensitivity limit of
our measurements, which is higher for lower luminosity objects
(see Sec.3.1),
 but also due to the
incompleteness of the BKA01 survey for
very low luminosity sources,  which are detected 
only when they have a large mid-IR excess, very likely
indicative of higher accretion rates. 
%All the objects with detected \Pab\ emission
%have a 6.7 \um\ excess $\simgreat 2$,
%while no object with smaller excess has \Pab\ emission.
%This suggests that probably the BKA01 sample is 
%incomplete below \Lstar$\sim 0.1$ \Lsun, rather than  at $\sim 0.03$ \Lsun,
%as estimated by the authors.

%\begin{figure}
%\begin{center}
%\leavevmode
%\centerline{ \psfig{file=ps.pab_eqwidth,width=9cm,angle=0} }
%\end{center}
%\caption{Equivalent width of \Pab as a function of \Lstar for
%Class II  (filled dots) and Class III objects (empty squares). Negative values
%are for lines in emission, positive for absorption.
%}
%\label{pab_width}
%\end{figure}

%\begin{figure}
%\begin{center}
%\leavevmode
%\centerline{ \psfig{file=ps.irdetected,width=9cm,angle=0} }
%\end{center}
%\caption{Distribution in luminosity of Class II objects with \Pab\ or
%\Brg\ emission (filled histogram). The empty histogram shows the
%distribution of the IR-observed sample.
%}
%\label{IR_detect}
%\end{figure}

\subsection {Accretion Luminosity}

\begin{figure}[ht!]
\begin{center}
\leavevmode
\centerline{ \psfig{file=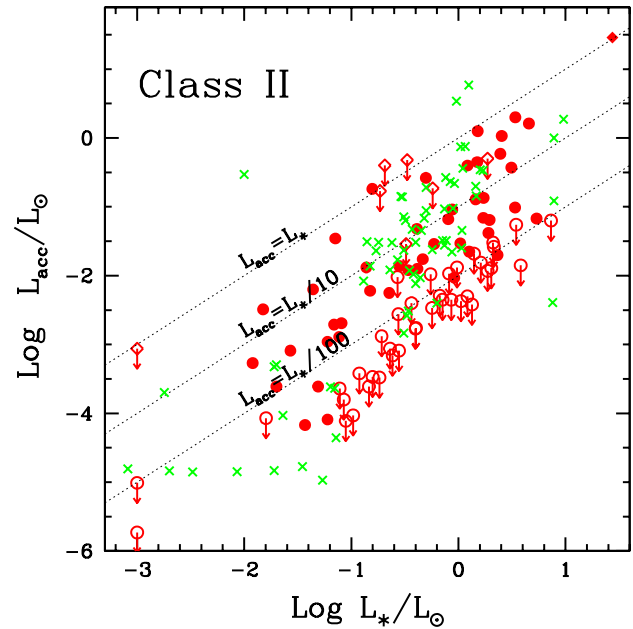,width=9cm,angle=0} }
%\centerline{ \psfig{file=ps.lacc_lstar,width=9cm,angle=0} }
\end{center}
\caption{
Accretion luminosity from the IR lines as function of \Lstar\ for
Class II objects. Dots show L$_{acc}$ measurements
from \Pab\ (filled:detections, empty: upper limits);
diamonds measurements from \Brg\ (filled: detections, empty: upper limits);
The dotted lines show the locus of
\Lacc/\Lstar=0.01, 0.1 and 1, as labelled. 
%Three objects with
%upper limits to L$_{acc}$ and \Lstar$< 0.01$ \Lsun\ are not shown.
Crosses are objects in Taurus
(see text for references).
}
\label{Lacc-Lstar}
\end{figure}

Fig.~\ref{Lacc-Lstar} shows the accretion luminosity of Class II
objects computed from the
IR line luminosity as a function of \Lstar. 
%First of all, note  that 
%our sensitivity limit on \Lacc\ increases strongly with \Lstar,
%this is because 
%our observations are basically limited in equivalent width
%(rather than in flux), to EW$\sim$0.3--1\AA\ for \Pab. 
%In our procedure,
%both the \Pab\ and the stellar luminosity are
%computed from the J-band stellar luminosity 
%(see Eq.2 and 3), and for a fixed value of EW one expects
%a relation \Lacc$\propto L_\star^{1.24}$.
%The observed distribution of \Lacc\ is only
%marginally steeper than that
%(\Lacc$\propto L_\star^{1.43\pm 0.1}$ for detections). 

For any given \Lstar, there is a large range of measured
\Lacc\ (about 50), which does not seem to vary significantly with
\Lstar; because of our sensitivity limit, this is probably just a lower
limit to the actual range of \Lacc. One can also see that for the
majority of objects \Lacc/\Lstar$<$0.1, but there is a significant fraction of 
cases with \Lacc$\sim$\Lstar.

%{\bf Comparison with Taurus: same}

\subsection {Mass accretion rate}

\begin{figure}[ht!]
\begin{center}
\leavevmode
\centerline{ \psfig{file=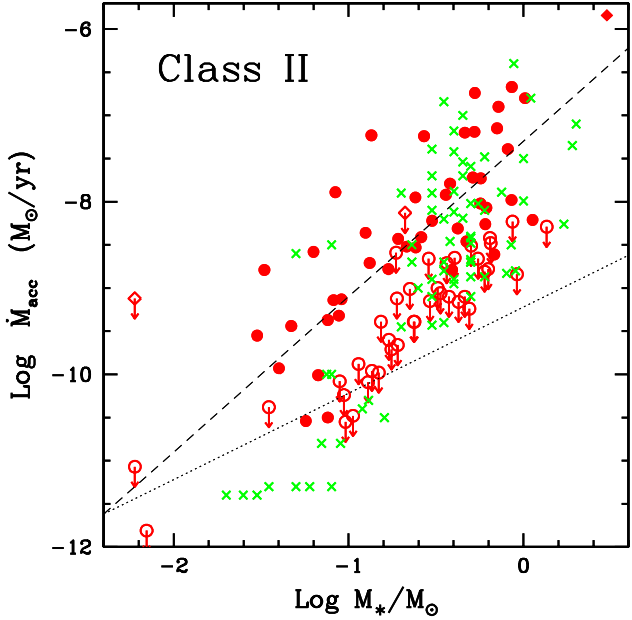,width=9cm,angle=0} }
%\centerline{ \psfig{file=ps.macc_mstar,width=9cm,angle=0} }
\end{center}
\caption{Mass accretion rate derived from the IR lines as function
of \Mstar. Symbols as in Fig.~\ref{Lacc-Lstar}. 
The  dashed line
shows the  relation \Macc$\propto$\Mstar$^{1.8}$,
derived from a statistical analysis using  ASURV;
the dotted line plots, for comparison, the relation
\Macc$\propto$\Mstar. 
}
\label{Macc-Mstar}
\end{figure}

Fig.~\ref{Macc-Mstar} shows the mass accretion rate \Macc\
of Class II sources
as function of \Mstar. 
There is a clear trend of increasing
\Macc\ with increasing \Mstar. Not including upper limits,
% (which are difficult to include because of their non-gaussian distribution)
we find using ASURV  (Feigelson and Nelson \cite{FN85})
\Macc $\propto$ \Mstar$^{1.8\pm 0.2}$; the  slope does not change
if we include the upper limits in the analysis.

Superimposed on this trend, there is a large spread of \Macc\
for any value of \Mstar, of two orders of magnitude at least.
Within statistical fluctuations, the objects are distributed quite
uniformly in this range.

%in the remaining of this section, we will
%examine  how robust it is against some of the assumptions we have used
%in its derivation.

%\begin{figure}[ht!]
%\begin{center}
%\leavevmode
%\centerline{ \psfig{file=ps.macc_lstar,width=9cm,angle=0} }
%\end{center}
%\caption{}
%\label{Macc-Lstar}
%\end{figure}

\section {Discussion}

The results summarized in Fig.~\ref{Macc-Mstar}
describe the accretion properties of the largest sample of
Class II stars in any single star-forming region
studied so far. The sample  contains more than hundred
objects with evidence of disks, and is complete in the mass interval
from  $\sim$0.03 to about 3 \Msun.  The corresponding
accretion rates  vary from $\sim 10^{-11}$ to $\sim 10^{-6}$ \Myr,
with a strong dependence of \Macc\ on \Mstar\ (\Macc$\propto
M_\star^{1.8\pm 0.2}$). For any \Mstar, there is a large dispersion
of values of \Macc, of two orders of magnitude at least, which  does
not seem to change with \Mstar. Note that the real spread
is likely bigger,
because of the many upper limits in our survey.
%the objects have the same age, we underestimate slightly the actual
%of the many upper limits we find in our survey.

\subsection {Ophiuchus and Taurus}

One of the  aims of our study was to compare the accretion properties
in Ophiuchus with those of objects in Taurus.
The Taurus results are shown by crosses in Fig.~\ref{Lacc-Lstar}
and \ref{Macc-Mstar}. 
The accretion luminosity
and mass accretion rate have been derived 
from the UV and optical
veiling and/or 
by fitting with magnetospheric accretion models
the \Ha\ profile. This second method is
the only possible one for very low mass objects and BDs, 
since veiling cannot be detected below a limiting value
\Macc$\simless 10^{-10}$ \Myr. The results are from
Gullbring et al.~(\cite{Gea98}),
Muzerolle et al.~(\cite{Mea98b}, \cite{Mea03}, \cite{Mea05}),
White \& Ghez (\cite{WG01}), White \& Basri (\cite{WB03}),
and Calvet et al.~(\cite{Cea04}); 
note that, for homogeneity, we  have re-determined \Mstar\ using DM98 tracks
for all objects.

The methods used to derive \Lacc\ and \Macc\ in the two regions
are therefore different, since
in Ophiuchus \Macc\ is derived from the luminosity
of the hydrogen recombination lines. However,
the relations
(eq.~3 and 4) we used  have been ``calibrated" mostly using Taurus
objects (see, e.g., Muzerolle et al.~\cite{Mea98b}, Calvet et al.~\cite{Cea04},
Natta et al.~\cite{Nea04}), so that we do not expect
any systematic difference in the Ophiuchus-Taurus comparison due to
the different methods.

The two figures show that the accretion properties of the
two star forming regions are very similar.
Muzerolle et al.~(\cite{Mea05}) derive 
\Macc $\propto$ \Mstar$^{2.1}$ for their sample (mostly in Taurus,
with additional brown dwarfs from other star-forming regions),
neglecting upper limits. Within the errors, this relation is 
identical to what we obtain in Ophiuchus.
If we concentrate on Fig.~\ref{Macc-Mstar}, we can see that not only
the slope of the relation of \Macc\ with \Mstar, but also
the range of values is very similar.
In particular, 
the two samples have similar values of the
maximum \Macc\ for any given \Mstar, and similar spread of \Macc\ values,
at least for \Mstar$\simgreat 0.06-0.08$ \Msun.

For lower \Mstar, most Taurus BDs have very low accretion rates,
1--2 orders of magnitude lower than similar objects in Ophiuchus.
As already discussed, the fact that we do not find these very low accretors in Ophiuchus
most likely reflects the incompleteness of the BAK01 sample at very
low masses,
and  selects 
objects with comparatively strong  mid-IR fluxes.
Natta et al.~(\cite{Nea02}) showed that the BAK01 sample of
brown dwarfs has relatively large luminosity, and is probably very
young. As discussed in Sec.~4.1, a 
fraction larger than for more luminous objects  has detected \Pab. 
All this indicates that there may be low \Macc\ BDs which are missing
from the Ophiuchus sample.
It is, in a way, more surprising that very few, if any,
of the  brown dwarfs in Taurus have  high \Macc,
while  higher mass objects in the  two regions have very similar
accretion properties.
It is possible that this difference between the two regions at
the very low end of the \Mstar\ distribution contains important
information, that needs further investigation.  This is, however,
beyond the scope of this paper.

%In the following, however, we will ignore it and
%summarize our results as follows. Both in Taurus and \Roph\
%\Macc\ is roughly  $\propto$\Mstar$^2$; for any given \Mstar,
%there is a range of measured \Macc\ of at least two order of magnitude.

\subsection {Variability}

All pre-main sequence stars are variable objects, and, in particular,
all the  accretion  indicators
in TTS and BDs show large variability.
%It is in general
%difficult to define   ``typical" accretion rates for any given star, unless
%long time-series of observations are available.

Variability does not affect
the correlation
of \Macc\ with \Mstar, 
as the \Roph\ sample is sufficiently large that
individual fluctuations  cannot change it. 
It  may be more important when we consider the spread
of \Macc\ values for any given \Mstar.
Recently, Scholz \& Jayawardhana (\cite{SJ05}) have 
studied the variability of accretion indicators (mostly H$_\alpha$) for
six young brown dwarfs; they claim that the accretion rate in some
of their objects varies by at least one order of magnitude, and that
this variability may account for the large spread in the \Macc -- \Mstar\
correlation.

We have  estimated the magnitude of the  spread in \Macc\
for individual objects by looking at the results of
Gatti et al.~(\cite{Gea06}), who have recently obtained J-band
spectra of  a small (14 objects)
subset of our Ophiuchus sample. The Gatti et al. sample includes
both TTS and BDs, observed  one to two years later than
the spectra discussed in this paper. The two data sets
show
variations in the \Pab\ equivalent width of a factor of two
at most (in both directions),
with only one exception, where the \Pab\ equivalent width
has increased by a factor of three over the
time interval  between the
two sets of observations. For the same
objects, we have also looked in the literature for 
variations of the 
broad-band J magnitude, used to compute
the line flux (Sec.3.2).
The variation of \Macc, computed taking the maximum variations in the J magnitude and in the \Pab\ equivalent width,
is of a factor $\sim 4$. This is much smaller
than the dispersion of points in Fig.~\ref{Macc-Mstar} and
would not change significantly any of our conclusions.

%For three  objects for which Gatti et al.~(\cite{Gea06})
%obtained K-band spectra,  three have been
%observed before (one  in this paper,
%two in LR99) and  we could compare different
%observations  of \Brg. The largest  difference in the equivalent width
%is a factor $\sim 4$; the resulting variation of \Macc\ (taking into
%account also the observed variability of the K-band magnitude) is
%roughly a factor $4$.

A detailed analysis of the variability of the IR emission lines and continuum,
in analogy to what has been done for H$_\alpha$  (e.g., Johns-Krull
\& Basri \cite{JKB97}),
is certainly needed. However, from the results obtained so far,
it seems unlikely that the dispersion of \Macc\ values  can be accounted
for by variability alone, and that, if averaged over a sufficiently long period
of time, one would find that
 all the
\Roph\ stars of a given mass accrete
at the same rate.

%However, we think that, even if Fig.~\ref{Macc-Mstar} is 
%only a snapshot and individual objects may have been observed
%when exceptionally quiet or exceptionally active,
%the spread of \Macc\ is very likely real. 

\subsection {Viscous disks}

The \Macc\ dependence on  \Mstar\
is difficult to understand in terms
of disk physics, as discussed, e.g., by Muzerolle et al.~(\cite{Mea03}),
Natta et al.~(\cite{Nea04}), Calvet et al.~(\cite{Cea04}). 
In a standard steady accretion disk model, \Macc\
is proportional to the disk mass divided by the time scale for
viscous evolution. In an $\alpha$-disk (Shakura \& Sunyaev \cite{SS73}),  
the viscosity depends 
on the ratio $\Omega/c_s^2$, where $\Omega$ is the keplerian angular velocity
and $c_s$ the sound speed; then,
 \Macc $\propto M_d \times M_\star^{-1/2} \times T_d$,
where $M_d$ and $T_d$ are disk mass and temperature, respectively. 
With the further assumptions that $M_d\propto M_\star$ (e.g., Natta et al.~\cite{Nea_PPIV}),
and that the disk heating is dominated by the stellar irradiation, this
gives, to zero order, \Macc $\propto M_\star^{1/2} \times T_\star$.
For PMS stars, the relation between \Tstar\ and \Mstar\ is
rather shallow (approximately
$T_\star \propto M_\star^{0.4}$ for \Mstar$\simgreat$0.1 \Msun,
and much flatter for lower masses; see, e.g., DM98) 
and we expect \Macc\ to increase
roughly as $M_\star^\gamma$, with $\gamma \simless 1$.
The relation will be even flatter if the contribution of the stellar radiation
to the disk heating is negligible.

It is possible that $\alpha$ (or, more generally, the efficiency
of momentum transfer) depends, in turn,  on \Mstar.
If viscosity is the result of magneto-rotational instabilities (MRI)
(see, e.g., Balbus \& Hawley \cite{BH91}), 
the disk gas should be sufficiently ionized.
Muzerolle et al.~(\cite{Mea03}) suggest that the steep correlation
of \Macc\ with \Mstar\ can be explained if 
the disk ionization is controlled by the X-ray radiation from the star, since
the X-ray luminosity  is not constant over
the mass spectrum, but is
observed to increase with \Mstar. 

X--ray observations of Ophiuchus have been recently carried out
with Chandra and XMM satellites by Imanishi et al. (\cite{ima_01})
and Ozawa et al. (\cite{ozawa}).
Both studies detected  a significant fraction of Class II sources
(70 and 48 \% respectively); they found that the X--ray spectral
properties, as well as the relationship between L$_{\rm X}$
and L$_{\rm bol}$  of class II sources are similar to those of class III
sources, but did not investigate the behaviour of X--ray luminosity
with stellar mass. To our knowledge, the only study addressing the
relation between mass and X-ray luminosity for young stars over a large
range of luminosities and masses is in Orion.
The COUP Chandra observations of
Orion  show that  $L_X$ scales
approximately as \Mstar$^{1.1-1.4}$
in the interval 0.1--2 \Msun\ (Preibisch et al.~\cite{Pea05}).
However, it is not clear that this variation of
$L_X$ is sufficient to produce the observed  \Macc--\Mstar correlation,
and more detailed MRI models, which include X-ray ionization,
 are required.
If the X-ray emission of the central star is controlling accretion,
the large spread of $L_X$ observed in the COUP data  could also
explain 
the large spread of \Macc\
for any given \Mstar.

Viscous disk models predict that \Macc\ decreases with time
(e.g.,  Hartmann et al.~\cite{Hea98}). 
Calvet et al.~(\cite{CHS_PPIV})
estimate  \Macc $\propto t^{-1.5}$, with a large uncertainty, from
a sample of TTS in Taurus, Chamaeleon and Ophiuchus. 
Neither the similarity of accretion rates between Ophiuchus and Taurus
nor the very large spread observed in both regions
support age as a main factor in the determination of
\Macc. 
If the Calvet et al.~(\cite{CHS_PPIV}) rate is correct, 
the difference in age between Taurus and Ophiucus should give on average
a difference in \Macc\ of a factor $\simgreat 3$, of which we have no evidence.
In addition,
the Ophiuchus \Macc\ range of more than two orders of magnitude
corresponds to an age range of at least a factor 20, much too large
when compared to the HR location of the objects (see, e.g., LR99).

%\subsection {Effects of companions}

The time evolution of viscous disks
is influenced by the presence of  close companions
(see Calvet et al.~\cite{CHS_PPIV}). Companions 
truncate the circumstellar disk at a radius which depends on the
binary separation. As the disk evolves, more and more matter expands outside
the truncation radius, with the effect of decreasing the disk mass and \Macc.
A sample of objects with the same initial  value of \Macc\
but companions at different distances will show with time an
increasing spread of \Macc\ values.

This effect, however, is not seen in the Taurus TTS
(White \& Ghez \cite{WG01}), where the accretion rate is similar for
single and primary stars with companions as close as 10 AU.
At the age of Ophiuchus, only very close
companions have had time to reduce \Macc\ by a significant
factor (separation $\simless$30 AU or $\simless 0.2$ arcsec
for an age of $10^6$ years
according to Calvet et al.~\cite{CHS_PPIV}).
There have been a number of multiplicity surveys of Ophiuchus,
some capable of detecting very close binaries. Three
Class II objects (i.e., objects with a mid-IR detected circumstellar disk)
have companions closer than
$\simless 0.25$ arcsec
(Barsony et al.~\cite{BKM03};
Ratzka et al.~\cite{Ratzka05}); 
one has  detected \Pab, while in the other two cases
the line has not been detected.
The observational evidence of a correlation between
the accretion rate and the presence of very close
companions is clearly inconclusive. At this stage, it cannot be
quantitatively confirmed nor dismissed, 
and should be investigated further.

%\begin{figure}
%\begin{center}
%\leavevmode
%\centerline{ \psfig{file=ps.macc_bin,width=9cm,angle=0} }
%\end{center}
%\caption{Mass accretion rate as function of \Mstar\ for
%single and wide  (separation $>2$ arcsec) binary
%stars (open squares) and close binaries (crosses). The plot includes
%all the Class II 
%which have been searched for companions. Arrows are upper limits.}
%\label{Macc-bin}
%\end{figure}

\subsection {Initial conditions}

Although all the effects discussed so far
can play a role and need further investigation,
it is possible that differences in the initial conditions,
i.e., in the physical properties of the molecular cores
from which the star+disk system forms, 
determine the TTS disk properties, and in particular the behaviour
of \Macc\ disussed in this paper.

The self-similar viscous disk  models of Hartmann et al.~(\cite{Hea98})
 show that the 
accretion rate is proportional to the disk mass  at t=0,
i.e., when  accretion onto the disk stops,
and, in the early phases ot the evolution,  to its t=0 outer radius,
which in turns depend on the core properties.
 Alexander \& Armitage (\cite{AA06}) have started  exploring
how this can introduce a \Macc $\propto$ \Mstar$^2$ correlation
at a later time.

More realistic models
that follow the formation and evolution of
circumstellar disks (Hueso \& Guillot \cite{HG05}) illustrate
clearly how different core properties (in particular, different
rotation velocities) can create a large
spread of \Macc\ for objects with the same \Mstar\ and age.
%how  cores of the same mass and different angular velocity,
%which accrete entirely onto the star+disk in a 0.15--0.18 My interval, 
%produce stars of the same mass  but very different disks, which
%will evolve differently in time. In the  two examples
%discussed in detail, a difference of a factor of ten in 
%the angular velocity
%(well within the observed range, see, e.g., Ohashi et al.~\cite{Oea97},
%Barranco \& Goodman \cite{BG98})
%and slightly different values of $\alpha$
%result in  a factor $\sim$6 difference in \Macc\ through the whole disk
%evolution.

Models that compute the  evolution of disks starting from the core
infall phase over a large range of parameters are required, if we
want to  estimate 
the effect of the initial conditions on the
relation of \Macc\ with \Mstar\ and on its scatter.
The observations presented in this paper, and the similar results
for Taurus, provide an excellent test of such models.
Note that the 
the fact that disk accretion properties in Taurus and Ophiuchus
are very similar, while the two regions have large
differences in their environment, should put strong constraints
on these models, which will be interesting to explore fully.

\section {Summary and conclusions}

In this paper, we report the results of a near-IR
spectroscopic survey of a large sample
of very young objects in the \Roph\ core.
The sample includes all
Class II objects, i.e., objects with evidence of circumstellar disks
from mid-IR photometry (BKA01). This sample
covers the  mass range between about 0.03 to 3 \Msun; according
to BKA01, it is
{\it complete} to a limiting magnitude of about 0.03 \Lsun,
or 0.05 \Msun. We have also observed   a significant fraction
of  Class III objects, i.e., with no mid-IR excess emission,
covering a similar range of luminosities.

In contrast to the Balmer lines,
the near-IR hydrogen recombination lines are seen in
emission only in Class II objects. Of all our Class III sample,
none has detected \Pab\ emission. This confirms our
assumption (Natta et al.~\cite{Nea04}) that the near-IR lines can provide
an immediate indication of the accreting properties of young stars,
even when only relatively low resolution spectra are available. 

We have derived  the mass accretion rate
\Macc\  from the luminosity of the hydrogen recombination
lines, mostly from \Pab\ but in few cases from \Brg. 
In total, we obtain  measurements of \Macc\ for 45 Class II objects, and upper limits for
39.

Our results show that 
\Macc\ increases sharply with \Mstar ($\propto M_\star^{1.8\pm0.2}$).
We also find a large range of values of \Macc\ for any given value of
\Mstar\ (a spread of roughly two order of
magnitudes,  independent of \Mstar). As discussed in the text,
this is likely a lower limit to the true dispersion.

When compared to accretion measurements in Taurus
(see Muzerolle et al.~\cite{Mea05} and references therein), we find that the two regions
look very similar, at least for objects with \Mstar$\simgreat 0.1$ \Msun.
For both Taurus and Ophiuchus, the dependence
of \Macc\ on \Mstar,  the upper envelope of the \Macc\ distribution (i.e., the
largest values of \Macc\ that any object of a given mass seems able to sustain),
and the range of \Macc\ values for any given \Mstar,
are very similar. At lower mass, the accretion rates of the Ophiuchus
objects are much larger than their Taurus analogs.

The observed behaviour of \Macc\ does not have an obvious explanation. 
The correlation of \Macc\ with \Mstar\ may be due to a dependence of
the disk physics on the properties of the central star.
Muzerolle et al.~(\cite{Mea03})
suggest as a cause
the effect of the X-ray emission from
the central star on the disk ionization and angular momentum transfer.
It is also possible that the correlation reflects 
the properties of the pre-stellar cores,
from which the star and disk form. Both possibilities need to be explored further.

The large spread of values of \Macc\ for any \Mstar\ may be a side-product
of the same mechanisms that produce the correlation between these
two quantities, as discussed in Sec.5. In addition, 
other effects may play an important role, for example the dynamical
action of close companions, or the intrinsic variability of the 
accretion process.

\begin{acknowledgements}
It is a pleasure to acknowledge the continuous, competent 
and friendly support of the ESO staff during the preparation 
and execution of the Visitor  and Service Mode observations at 
La Silla and Paranal observatories.
We whish to thank an anonymous referee for very useful comments.
This project was partially supported by MIUR grants 2002028843/2002
and 2004025227/2004.
\end{acknowledgements}

%\begin{appendix} 
\appendix
\section{Testing the assumption of coeval star formation}

The assumption of coeval star formation, albeit quite reasonable
for a  region like \Roph, introduces errors in our results.
The same is true of the choice of any specific set of evolutionary tracks.
However, it turns out  that both kinds of errors are unimportant,
when dealing with
a large sample of objects as in our case.

\begin{figure}[ht!]
\begin{center}
\leavevmode
\centerline{ \psfig{file=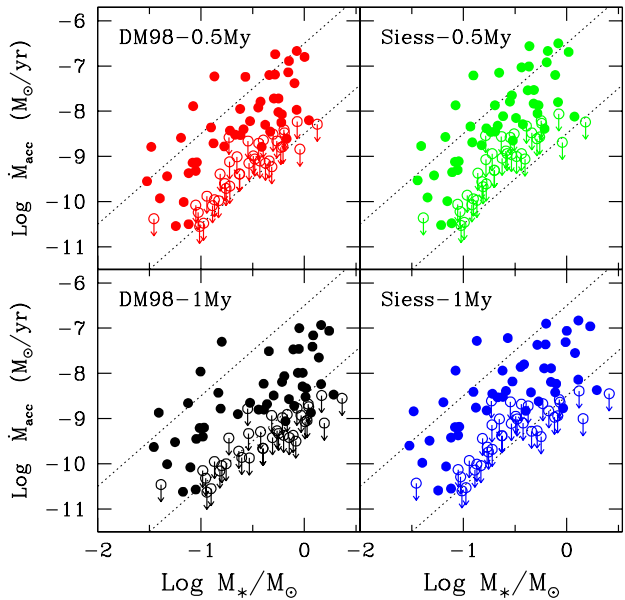,width=9cm,angle=0} }
%\centerline{ \psfig{file=ps.macc_mstar_tracks,width=9cm,angle=0} }
\end{center}
\caption{Same as Fig.~\ref{Macc-Mstar} for different ages and evolutionary
tracks. The top-left panel is for DM98 0.5My (as in Fig.~\ref{Macc-Mstar}),
the bottom-left for DM98, 1 My, the top right is for Siess (\cite{Siess00})\
 evolutionary
tracks at 0.5 My, the bottom right  at 1 My. In each Panel, the 
two dotted lines (\Macc $\propto$ \Mstar$^2$) have been drawn to guide the eye
in the comparison.}
\label{Macc_tracks}
\end{figure}

Fig.~\ref{Macc_tracks} shows the analog of Fig.~\ref{Macc-Mstar},
reproduced on the top-left panel,
computed using the DM98 isochrone for 1 My and the evolutionary 
tracks of Siess (\cite{Siess00}) for 0.5 My and 1 My, respectively.
Older tracks give slightly lower values of \Macc,
especially for more massive objects,
while the range of \Mstar\ remains practically
the same. Adopting different evolutionary
tracks does not change the results.
The main consequence  of assuming coeval star formation
is to reduce slightly the real spread of \Macc\ for any
given value of \Mstar.

%Using the 1 My isochrone DM98 or the evolutionary tracks
%of Siess for 0.5 My and 1 My show that, for a fixed \Lstar,
%\Mstar\  has a maximum variation of 0.25 dex, and the
%ratio \mr\ of roughly 0.5 dex; the diffences are maximum around
%$\sim$1 \Lsun\ and, obviously, older isochrones give larger
%values of both \Mstar\ and \mr.

\section{Taurus: a test of the method}

A validation of the  method used to compute the two
quantities \Macc\ and \Mstar
and an estimate of the errors can be obtained by
applying the same procedure to a sample of objects with known stellar
parameters and accretion rates.
% determined not from
%the \Pab\ luminosity but with other, independent 
%methods (i.e., from veiling and by fitting \Ha\ profiles).

The only  sample for which this is possible is Taurus,
which has been studied extensively over a large range of masses
We have taken all the Taurus objects for which we could find in the
literature reliable stellar parameters and accretion rates,
measured from veiling and/or by fitting the observed \Ha\ profiles
with magnetospheric accretion models
(Muzerolle et al.~\cite{Mea98a}, \cite{Mea03}, \cite{Mea05},
Calvet et al.~\cite{Cea04}).
For those with published \Pab\ fluxes or equivalent widths,
we  have  followed the same procedure used for the \Roph\
sample. We have first computed \Lacc\ from L(\Pab), and determined
the stellar parameters \mr\ and \Mstar\ from \Lstar,  assuming coeval
star formation at 1 My and the DM98 evolutionary tracks.
\Macc\ is then derived from \Lacc\ and \mr.

\begin{figure}[ht!]
\begin{center}
\leavevmode
\centerline{ \psfig{file=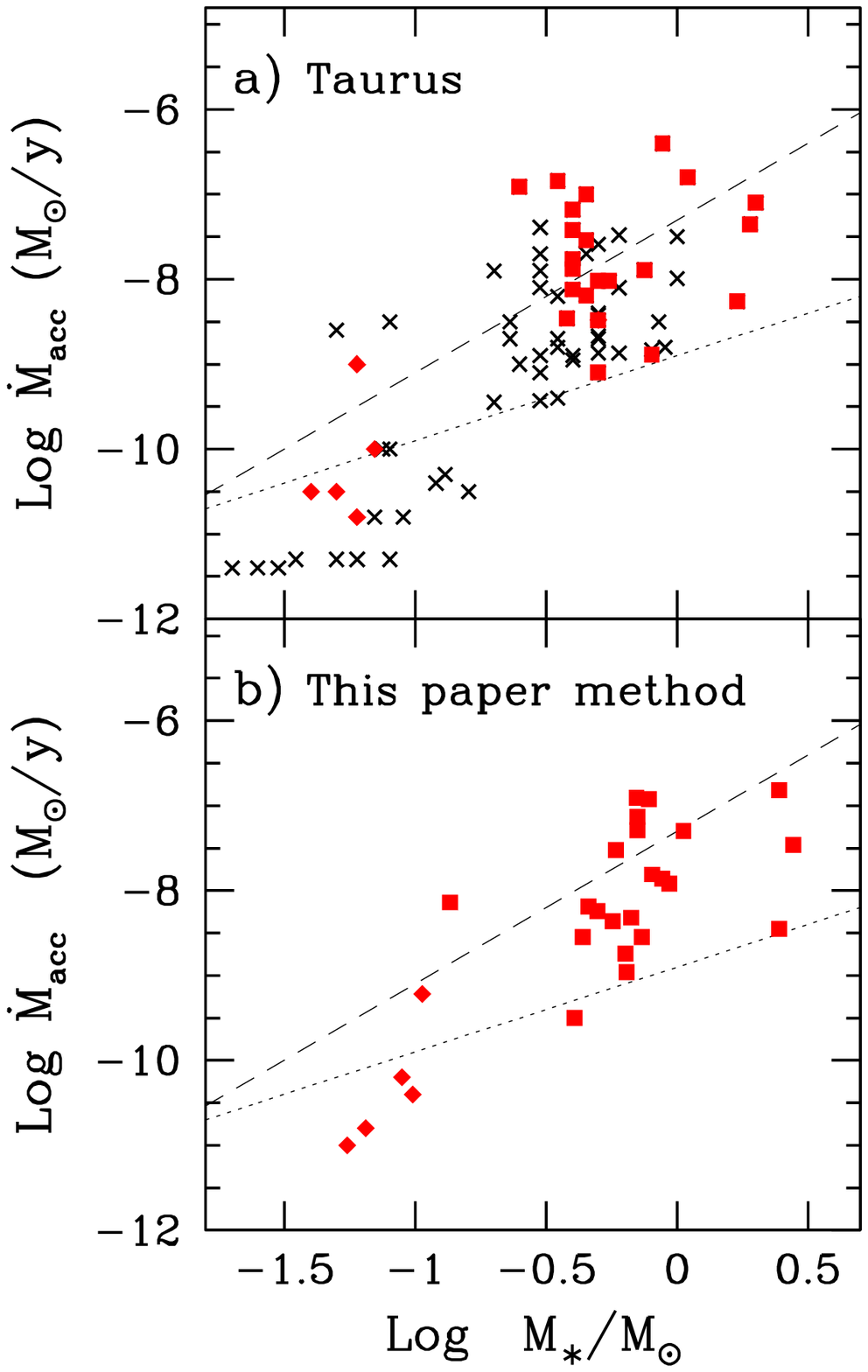,width=9cm,angle=0} }
%\centerline{ \psfig{file=ps.macc_mstar_taurus,width=9cm,angle=0} }
% ricordarsi di cambiare i bounding box -->18 144 380 718
\end{center}
\caption{Top Panel: \Macc\ vs. \Mstar\ for objects in Taurus for which
\Macc\ has been obtained from veiling and/or by fitting the \Ha\
profiles with magnetospheric accretion models (crosses and squares).
Data from Gullbring et al.~(\cite{Gea98}), Muzerolle et al.~(\cite{Mea98b}, 
\cite{Mea03}, \cite{Mea05}),
White \& Ghez (\cite{WG01}), 
White \& Basri (\cite{WB03}), Calvet et al.~(\cite{Cea04}). Squares identify
objects for which we could find in the literature \Pab\ observations.
Diamonds are very low mass objects in Ophiuchus and Chamaeleon with
\Macc\ estimates from \Ha\ profile fitting (from Natta et al.~\cite{Nea04}).
Bottom panel: both \Mstar\ and \Macc\ have been derived
following the method we use for the Ophiuchus objects in this paper
(see text). As in Fig.~\ref{Macc-Mstar}, the two dotted lines show the run
of the \Macc $\propto$ \Mstar$^{1.8}$ and \Macc $\propto$ \Mstar relationships.
}
\label{taurus}
\end{figure}

The results are summarized in the Fig.~\ref{taurus}.
The top panel shows the complete sample of Taurus objects for which we could
find measurements of \Macc\ in the literature. The  squares are those
for which also \Pab\ data exist; because none of the BDs in Taurus
has a published J-band spectrum, we have added the BDs in Ophiuchus
and Chamaeleon for which Natta et al.~(\cite{Nea04}) have measured \Macc\ from
model fitting of the \Ha\ profiles.
The bottom panel shows the same plot when both \Mstar\ and \Macc\ are derived
from the observed \Lstar\ and \Pab\ luminosity, as  for the \Roph\ stars.

The results indicate that our procedure does not introduce systematic
trends in the results. The trend of \Macc\ increasing sharply
with \Mstar\ is reproduced  in our method,
and also the range of \Macc\ for a given
\Mstar\ is similar, even if, as expected, the assumption of
coeval star formation underestimates its spread slightly. 

\section{Tables}
Tables are available in electronic form only.

%\end{document}

\newpage
\pagestyle {empty}
\topmargin 5cm

{\small
\begin{landscape}
\begin{longtable}{clccccccccccccccl}
\caption{ Class II Objects}\\

\hline\hline
\hline\hline

(1)   & (2)   & (3)& (4)  & (5) & (6) & (7)& (8) & (9)& (10)& (11)& (12)& (13)& (14)& (15)& (16) &(17)\\ 
\# &Object&\multicolumn{2}{c} {Coordinates}& J& H& K& A$_{\rm J}$  &  Lg L$_\ast$& Lg T$_{eff}$& Lg M$_\ast$& EW (Pa$_\beta$)&Inst. & L(Pa$_\beta$)& L$_{acc}$& $\dot M_{acc}$& Other \\
&(ISO\#)& \multicolumn{2}{c}{(J2000.0)}& \multicolumn{3}{c}{(mag)} & (mag)& (L$_\odot$)& (K)& (M$_\odot$)& (\AA)& & (L$_\odot$)& (L$_\odot$)& (M$_\odot$/y)&  Names \\

\hline
\endfirsthead
\caption{continued.}\\
\hline\hline
\endhead
\endfoot

  1 &$\rho$Oph-ISO 001   & 16 25 36.74& -24 15 42.40&  10.42&  9.04&  8.38&     1.8&  0.21&  3.59& -0.26& $<$-0.7&S& $<$-4.27& $<$-1.81& $<$-8.66&    IRS2                \\
  2 &$\rho$Oph-ISO 002$^+$  & 16 25 38.12& -24 22 36.30&  12.84& 10.75&  9.54&     3.1& -0.23&  3.53& -0.52&    -2.8&S&   -4.07&   -1.54&    -8.22&    B162538-242238      \\
  3 &$\rho$Oph-ISO 003   & 16 25 39.58& -24 26 34.90&  11.89& 10.05&  8.95&     2.5& -0.10&  3.55& -0.44&    -3.9&S&   -3.81&   -1.18&    -7.92&    IRS3                \\
  4 &$\rho$Oph-ISO 006   & 16 25 56.16& -24 20 48.20&   9.15&  8.14&  7.52&     0.5&  0.18&  3.59& -0.28&   -19.0& S&  -2.87&    0.10&    -6.74&    SR4/IRS12    \\
  5 &$\rho$Oph-ISO 009   & 16 26  1.37& -24 25 20.40&  14.43& 12.44& 11.24&     2.8& -1.07&  3.45& -1.03& $<$-0.3&I& $<$-5.74& $<$-3.80& $<$-10.24&    SKS1-4              \\
  6 &$\rho$Oph-ISO 012   & 16 26  4.58& -24 17 51.50&  15.79& 13.41& 12.19&     4.2& -1.05&  3.45& -1.02& $<$-0.2&I& $<$-5.96& $<$-4.11& $<$-10.55&    B162604-241753      \\
  7 &$\rho$Oph-ISO 013   & 16 26  7.04& -24 27 24.20&  15.35& 12.38& 10.64&     5.2& -0.38&  3.51& -0.62&    -2.1&S&   -4.34&   -1.90&    -8.53&    B162607-242725      \\
  8 &$\rho$Oph-ISO 017   & 16 26 10.33& -24 20 54.80&  14.37& 10.85&  8.47&     5.8&  0.33&  3.61& -0.19& $<$-0.8& S&$<$-4.11& $<$-1.58& $<$-8.48&    GSS26               \\
 9 &$\rho$Oph-ISO 019   & 16 26 16.84& -24 22 23.20&  11.03&  9.13&  8.20&     3.1&  0.54&  3.64& -0.06& $<$-0.9&S& $<$-3.87& $<$-1.26& $<$-8.23&    GSS29/EL18          \\
 10 &$\rho$Oph-ISO 020   & 16 26 17.06& -24 20 21.60&   9.65&  8.61&  8.06&     0.8&  0.10&  3.58& -0.33&    -1.2&S&   -4.15&   -1.65&    -8.46&    DoAr24/GSS28        \\
 11 &$\rho$Oph-ISO 023   & 16 26 18.82& -24 26 10.50&  14.84& 13.20& 12.14&     1.8& -1.69&  3.42& -1.40&    -1.8& I&  -5.59&   -3.61&    -9.93&    SKS1-BDN04          \\
 12 &$\rho$Oph-ISO 024   & 16 26 18.87& -24 28 19.70&  12.58&  9.93&  8.07&     3.7&  0.18&  3.59& -0.28&    -8.9& I&  -3.20&   -0.35&    -7.19&    VSSG1               \\
 13 &$\rho$Oph-ISO 026  & 16 26 20.97& -24  8 51.90&  10.88&  9.87&  9.50&     1.0& -0.33&  3.52& -0.59&    -2.4& I&  -4.23&   -1.76&    -8.41&    RBR15        \\
 14 &$\rho$Oph-ISO 030   & 16 26 21.53& -24 26  1.00&  12.57& 11.52& 10.92&     0.7& -1.22&  3.44& -1.12&    -0.3& I&  -5.95&   -4.09&   -10.50&    GY5                 \\
 15 &$\rho$Oph-ISO 032   & 16 26 21.90& -24 44 39.80&  12.34& 11.48& 10.86&     0.0& -1.43&  3.43& -1.24&    -0.4&I&   -6.01&   -4.17&   -10.54&    GY3                 \\
 16 &$\rho$Oph-ISO 033   & 16 26 22.27& -24 24  7.10&  16.45& 15.09& 13.94&     0.6& -2.95&  3.37& -2.16& $<$-0.7& I&$<$-7.15& $<$-5.73& $<$-11.81&    GY11                \\
 17 &$\rho$Oph-ISO 035   & 16 26 22.96& -24 28 46.10&  14.93& 12.80& 11.53&     3.2& -1.12&  3.45& -1.06&    +0.6& I&  --&  --&   --&    GY15                \\
 18 &$\rho$Oph-ISO 036   & 16 26 23.36& -24 20 59.80&   8.97&  7.50&  6.57&     1.5&  0.73&  3.66&  0.05&    -0.7& S&  -3.80&   -1.17&    -8.21&    GSS31/GY20A         \\
 19 &$\rho$Oph-ISO 037   & 16 26 23.58& -24 24 39.50&  15.05& 12.25& 10.22&     3.9& -0.82&  3.46& -0.88&    -3.0& S&  -4.58&   -2.22&    -8.71&    LFAM3/GY21          \\
 20 &$\rho$Oph-ISO 038   & 16 26 23.68& -24 43 13.90&   9.39&  8.40&  7.85&     0.6&  0.13&  3.58& -0.31& $<$-0.3&S& $<$-4.72& $<$-2.42& $<$-9.24&    DoAr25/GY17         \\
 21 &$\rho$Oph-ISO 039   & 16 26 24.04& -24 24 48.10&  11.12&  8.72&  7.32&     3.9&  0.86&  3.68&  0.13& $<$-0.5&S& $<$-3.83& $<$-1.20& $<$-8.29&    S2/GY23             \\
 22 &$\rho$Oph-ISO 040   & 16 26 24.07& -24 16 13.50&  10.00&  8.09&  6.68&     2.0&  0.53&  3.63& -0.07&   -12.7& S&  -2.72&    0.30&    -6.67&    EL24                \\
 23 &$\rho$Oph-ISO 041   & 16 26 25.28& -24 24 45.00&  16.28& 13.10& 11.07&     5.3& -0.74&  3.46& -0.83& $<$-0.3&I& $<$-5.50& $<$-3.48& $<$-9.98&    GY29       \\
 24 &$\rho$Oph-ISO 043   & 16 26 27.54& -24 41 53.50&  14.04& 11.42&  9.98&     4.6& -0.09&  3.55& -0.44& $<$-1.0&S& $<$-4.39& $<$-1.97& $<$-8.71&    GY33                \\
 25 &$\rho$Oph-ISO 046   & 16 26 30.47& -24 22 57.10&  16.31& 12.55&  9.98&     6.3& -0.30&  3.52& -0.57&   -16.5&I&   -3.37&   -0.58&    -7.24&    VSSG27/GY51         \\
 26 &$\rho$Oph-ISO 051   & 16 26 36.83& -24 15 51.90&  12.66& 10.83&  9.59&     2.1& -0.61&  3.48& -0.75& $<$-0.4& S&$<$-5.27& $<$-3.16& $<$-9.71&    B162636-241554      \\
 27 &$\rho$Oph-ISO 052   & 16 26 37.79& -24 23  0.70&  15.74& 12.91& 11.11&     4.5& -0.86&  3.46& -0.90&    -5.9& I&  -4.32&   -1.88&    -8.36&    VSSG4/GY81          \\
 28 &$\rho$Oph-ISO 053   & 16 26 38.60& -24 23 10.00&  15.29& 12.89& 11.63&     4.2& -0.84&  3.46& -0.89& $<$-0.3&I& $<$-5.59& $<$-3.61& $<$-10.09&    GY84                \\
 29 &$\rho$Oph-ISO 056   & 16 26 41.26& -24 40 18.00&  10.77&  9.77&  9.27&     0.8& -0.40&  3.51& -0.63& $<$-0.5&S& $<$-4.98& $<$-2.77& $<$-9.39&    WSB37/GY93          \\
 30 &$\rho$Oph-ISO 062   & 16 26 42.86& -24 20 29.90&  10.50&  8.77&  7.88&     2.5&  0.53&  3.63& -0.07&    -1.4& S&  -3.68&   -1.01&    -7.98&    GSS37/GY110         \\
 31 &$\rho$Oph-ISO 063   & 16 26 42.89& -24 22 59.10&  15.33& 12.82& 11.44&     4.3& -0.80&  3.46& -0.87& $<$-0.3&I& $<$-5.50& $<$-3.47& $<$-9.96&    GY109               \\
 32 &$\rho$Oph-ISO 067   & 16 26 45.03& -24 23  7.70&  13.25& 10.60&  8.96&     4.2&  0.09&  3.57& -0.33&   -10.0&S&   -3.23&   -0.40&    -7.20&    GSS39/GY116         \\
 33 &$\rho$Oph-ISO 068$^+$  & 16 26 46.43& -24 12  0.10&   9.68&  8.31&  7.49&     1.4&  0.37&  3.61& -0.17&    -0.6& S&  -4.19&   -1.70&    -8.61&    VSS27               \\
 34 &$\rho$Oph-ISO 072$^+$  & 16 26 48.98& -24 38 25.20&  13.50& 11.44&  9.98&     2.5& -0.81&  3.46& -0.87&   -36.0& S&  -3.49&   -0.74&    -7.23&    WL18/GY129          \\
 35 &$\rho$Oph-ISO 078   & 16 26 54.44& -24 26 20.70&  14.70& 11.69& 10.01&     5.5&  0.02&  3.57& -0.37& $<$-0.4&I& $<$-4.69& $<$-2.37& $<$-9.16&    VSSG5/GY153         \\
 36 &$\rho$Oph-ISO 079   & 16 26 54.77& -24 27  2.20&   0.00& 14.87& 12.87&     6.0& -1.24&  3.44& -1.13& $<$-3.0&I& --& --& --&    GY154               \\
 37 &$\rho$Oph-ISO 083   & 16 26 56.66& -24 13 53.80&  12.26& 10.31&  9.25&     3.0& -0.06&  3.56& -0.42&    -4.5&S&   -3.71&   -1.04&    -7.79&    B162656-241353      \\
 38 &$\rho$Oph-ISO 084   & 16 26 57.33& -24 35 38.80&   0.00& 15.09& 12.81&     7.1& -0.99&  3.45& -0.98& $<$-5.0&I& --& --& --&    WL21/GY164          \\
 39 &$\rho$Oph-ISO 086   & 16 26 58.40& -24 21 30.00&  16.01& 13.11& 11.46&     5.1& -0.72&  3.47& -0.82& $<$-0.8&S& $<$-5.06& $<$-2.88& $<$-9.39&    IRS26/GY171         \\
 40 &$\rho$Oph-ISO 087   & 16 26 58.64& -24 18 34.70&  15.45& 13.01& 11.48&     3.7& -1.09&  3.45& -1.04&    -2.4& I&  -4.92&   -2.69&    -9.13&    B162658-241836      \\
 41 &$\rho$Oph-ISO 088a  & 16 26 58.51& -24 45 36.90&   9.75&  8.16&  7.06&     1.5&  0.40&  3.62& -0.14&   -10.5&S&   -2.92&    0.03&    -6.90&    SR24N/GY168         \\
 42 &$\rho$Oph-ISO 088b  & 16 26 58.44& -24 45 31.90&  10.37&  8.63&  7.55&     2.1&  0.39 &  3.62& -0.15&    -7.0&S&   -3.11&   -0.23&    -7.15&    SR24S/GY167 \\ 
43 &$\rho$Oph-ISO 089   & 16 26 59.05& -24 35 56.90&  16.05& 13.35& 11.82&     4.7& -0.93&  3.46& -0.94& $<$-0.5&I& $<$-5.45& $<$-3.42& $<$-9.88&    WL14/GY172          \\
 44 &$\rho$Oph-ISO 092   & 16 27  2.34& -24 37 27.20&  14.16& 10.48&  8.06&     6.3&  0.66&  3.65&  0.01&    -8.4& S&  -2.79&    0.21&    -6.80&    WL16/GY182      \\
 45 &$\rho$Oph-ISO 093   & 16 27  3.01& -24 26 14.70&   0.00& 15.65& 12.56&    10.6& -0.24&  3.53& -0.53& $<$-20.0&S& --& --& --&    GY188  \\
 46 &$\rho$Oph-ISO 094   & 16 27  3.59& -24 20  5.40&  17.24& 14.91& 13.56&     3.7& -1.90&  3.41& -1.53&    -5.0& I&  -5.34&   -3.27&    -9.55&    B162703-242007      \\
 47 &$\rho$Oph-ISO 095   & 16 27  4.11& -24 28 29.90&  16.90& 13.09& 10.86&     7.2& -0.15&  3.54& -0.48& $<$-0.6&I& $<$-4.67& $<$-2.35& $<$-9.06&    WL1/GY192           \\
 48 &$\rho$Oph-ISO 098   & 16 27  4.57& -24 27 15.70&  16.48& 13.03& 11.22&     6.8& -0.17&  3.54& -0.49& $<$-0.7&I& $<$-4.62& $<$-2.29& $<$-9.00&    GY195               \\
 49 &$\rho$Oph-ISO 102   & 16 27  6.60& -24 41 48.80&  12.43& 11.40& 10.77&     0.6& -1.22&  3.44& -1.12&    -2.0& I&  -5.12&   -2.96&    -9.37&    GY204               \\
 50 &$\rho$Oph-ISO 103   & 16 27  6.78& -24 38 15.00&   0.00& 14.30& 10.97&    11.6&  0.71&  3.66&  0.04& $<$-0.6&I& --& --& --&    WL17/GY205          \\
 51 &$\rho$Oph-ISO 105   & 16 27  9.10& -24 34  8.10&  12.55& 10.19&  8.91&     4.0&  0.29&  3.60& -0.21&    -1.7& I&  -3.82&   -1.19&    -8.07&    WL10/GY211          \\
 52 &$\rho$Oph-ISO 106   & 16 27  9.07& -24 12  0.80&  12.41& 10.73&  9.80&     2.3& -0.44&  3.50& -0.65& $<$-1.0&S& $<$-4.71& $<$-2.40& $<$-9.01&    B162708--241204     \\
 53 &$\rho$Oph-ISO 107   & 16 27  9.35& -24 40 22.40&   0.00& 13.55& 11.30&     7.0& -0.30&  3.52& -0.56& $<$-1.3&I& --& --& --&    GY213               \\
 54 &$\rho$Oph-ISO 110   & 16 27 10.28& -24 19 12.70&   8.74&  7.51&  6.72&     0.9&  0.58&  3.64& -0.04& $<$-0.3&I& $<$-4.30& $<$-1.85& $<$-8.84&    SR21/VSSG23  \\
 55 &$\rho$Oph-ISO 112   & 16 27 11.18& -24 40 46.70&   0.00& 12.88& 10.20&     8.9&  0.57&  3.64& -0.04&   -17.9& I&  --&  --&   --&    GY224               \\
 56 &$\rho$Oph-ISO 115   & 16 27 12.13& -24 34 49.10&  15.62& 13.11& 11.49&     3.7& -1.16&  3.45& -1.08&    -2.7& I&  -4.94&   -2.71&    -9.14&    WL11/GY229          \\
 57 &$\rho$Oph-ISO 116   & 16 27 13.73& -24 18 16.90&  12.26& 10.25&  9.29&     3.4&  0.15&  3.58& -0.30& $<$-1.0&S& $<$-4.18& $<$-1.68& $<$-8.51&    B162713-241818      \\
 58 &$\rho$Oph-ISO 117   & 16 27 13.82& -24 43 31.70&  13.32& 11.23&  9.98&     3.1& -0.47&  3.50& -0.67&    -2.4& S&  -4.36&   -1.92&    -8.52&    GY235               \\
 59 &$\rho$Oph-ISO 118   & 16 27 14.51& -24 26 46.10&   0.00& 15.34& 12.26&    10.5& -0.11&  3.55& -0.45& $<$-20.0&I& --& --& --&    IRS33/GY236         \\
 60 &$\rho$Oph-ISO 120   & 16 27 15.45& -24 26 39.80&  17.42& 13.46& 10.79&     6.8& -0.57&  3.49& -0.73& $<$-2.5& I&$<$-4.43& $<$-2.02& $<$-8.59&    IRS34/GY239         \\
 61 &$\rho$Oph-ISO 121a  & 16 27 15.88& -24 38 43.40&  13.89& 11.26&  9.59&     4.1& -0.25&  3.53& -0.53& $<$-0.6& S&$<$-4.76& $<$-2.47& $<$-9.15&    WL20/GY240A         \\
 62 &$\rho$Oph-ISO 121b  & 16 27 15.70& -24 38 43.40&  13.57& 10.87&  9.48&     5.0&  0.31 &  3.60& -0.20& $<$-0.5&S& $<$-4.33& $<$-1.89& $<$-8.78&    WL20/GY240B    \\
 63 &$\rho$Oph-ISO 123   & 16 27 17.59& -24  5 13.70&  12.73& 11.49& 10.73&     1.0& -1.15&  3.45& -1.08&   -21.9&I&   -4.02&   -1.46&    -7.89&    ISO 1627176-240519   \\
 64 &$\rho$Oph-ISO 124   & 16 27 17.57& -24 28 56.30&   0.00& 14.42& 11.58&     9.5&  0.03&  3.57& -0.37& $<$-10.0&I& --& --& --&    IRS37/GY244         \\
 65 &$\rho$Oph-ISO 128   & 16 27 18.49& -24 29  5.90&  14.61& 11.50&  9.68&     5.5&  0.08&  3.57& -0.34& $<$-0.4&S& $<$-4.63& $<$-2.30& $<$-9.10&    WL4/GY247           \\
 66 &$\rho$Oph-ISO 129   & 16 27 19.22& -24 28 43.90&   0.00& 14.66& 11.49&    10.9&  0.33&  3.61& -0.19& $<$-30.0&I& --& --& --&    WL3/GY249           \\
 67 &$\rho$Oph-ISO 132   & 16 27 21.47& -24 41 43.10&  15.22& 11.25&  8.48&     6.6&  0.32&  3.61& -0.19& $<$-0.9&S& $<$-4.06& $<$-1.52& $<$-8.42&    IRS42/GY252         \\
 68 &$\rho$Oph-ISO 138   & 16 27 26.22& -24 19 23.00&  16.40& 14.24& 12.93&     3.1& -1.79&  3.42& -1.46& $<$-1.0&I& $<$-5.94& $<$-4.07& $<$-10.38&    B162726-241925      \\
 69 &$\rho$Oph-ISO 140   & 16 27 26.49& -24 39 23.10&  15.69& 12.07&  9.95&     6.7&  0.16&  3.58& -0.29&    -3.7& S&  -3.60&   -0.89&    -7.72&    GY262               \\
 70 &$\rho$Oph-ISO 142   & 16 27 27.38& -24 31 16.60&  12.35& 10.38&  9.32&     3.0& -0.07&  3.55& -0.43& $<$-0.5&S& $<$-4.67& $<$-2.35& $<$-9.10&    VSSG25/GY267        \\
 71 &$\rho$Oph-ISO 144   & 16 27 28.45& -24 27 21.00&  15.74& 12.31& 10.10&     5.8& -0.26&  3.53& -0.54& $<$-1.4&I& $<$-4.40& $<$-1.98& $<$-8.66&    IRS45/GY273         \\
 72 &$\rho$Oph-ISO 147   & 16 27 30.18& -24 27 43.40&  15.32& 11.52&  9.02&     6.6&  0.27&  3.60& -0.22& $<$-0.5&S& $<$-4.36& $<$-1.93& $<$-8.81&    IRS47/GY279         \\
 73 &$\rho$Oph-ISO 151   & 16 27 30.84& -24 24 56.00&  12.70& 10.95& 10.07&     2.6& -0.40&  3.51& -0.62& $<$-0.5&S& $<$-4.97& $<$-2.76& $<$-9.39&    GY284               \\
 74 &$\rho$Oph-ISO 154   & 16 27 32.85& -24 32 34.80&  16.19& 12.74& 10.96&     6.9& -0.01&  3.56& -0.39& $<$-1.0&S& $<$-4.32& $<$-1.88& $<$-8.65&    GY291               \\
 75 &$\rho$Oph-ISO 155   & 16 27 33.11& -24 41 15.30&  11.32&  9.13&  7.81&     3.2&  0.49&  3.63& -0.09&    -4.0& S&  -3.26&   -0.43&    -7.39&    GY292               \\
 76 &$\rho$Oph-ISO 160   & 16 27 37.42& -24 17 54.90&  14.15& 12.76& 11.95&     1.5& -1.57&  3.43& -1.33&    -3.3&I&   -5.22&   -3.09&    -9.44&    B162737-241756      \\
 77 &$\rho$Oph-ISO 163   & 16 27 38.32& -24 36 58.60&  11.38&  9.43&  8.27&     2.7&  0.23&  3.59& -0.25&    -2.0& S&  -3.80&   -1.16&    -8.02&    IRS49/GY308         \\
 78 &$\rho$Oph-ISO 164   & 16 27 38.63& -24 38 39.20&  13.27& 11.93& 11.08&     1.2& -1.31&  3.44& -1.17&    -0.8&I&   -5.60&   -3.61&   -10.01&    GY310               \\
 79 &$\rho$Oph-ISO 165   & 16 27 38.94& -24 40 20.70&  16.54& 13.91& 12.29&     4.2& -1.35&  3.44& -1.20&    -9.6&I&   -4.56&   -2.20&    -8.58&    GY312               \\
 80 &$\rho$Oph-ISO 166   & 16 27 39.43& -24 39 15.50&  10.75&  9.21&  8.46&     2.2&  0.24&  3.59& -0.25&    -3.3& I&  -3.58&   -0.87&    -7.73&    GY314               \\
 81 &$\rho$Oph-ISO 168   & 16 27 40.29& -24 22  4.00&   8.44&  7.67&  7.21&     0.0&  0.28&  3.60& -0.22&    -1.3&S&   -3.96&   -1.38&    -8.26&    SR9/GY319/IRS52     \\
 82 &$\rho$Oph-ISO 170   & 16 27 41.61& -24 46 44.70&  17.20& 15.33& 13.55&     1.0& -3.07&  3.36& -2.23& $<$-3.0&I& $<$-6.63& $<$-5.01& $<$-11.07&    B162741-244645      \\
 83 &$\rho$Oph-ISO 171   & 16 27 41.75& -24 43 36.10&   0.00& 14.88& 12.29&     8.5& -0.50&  3.50& -0.68& $<$-4.0&I& --& --& --&    GY323               \\
 84 &$\rho$Oph-ISO 172   & 16 27 42.70& -24 38 50.60&  13.24& 11.44& 10.54&     2.8& -0.56&  3.49& -0.72& $<$-1.0&S& $<$-4.82& $<$-2.56& $<$-9.12&    GY326               \\
 85 &$\rho$Oph-ISO 175   & 16 27 45.79& -24 44 53.90&  17.38& 14.54& 12.46&     4.0& -1.82&  3.42& -1.48&   -15.7& I&  -4.77&   -2.49&    -8.79&    GY344               \\
 86 &$\rho$Oph-ISO 176   & 16 27 46.29& -24 31 41.20&  13.83& 12.21& 11.32&     2.2& -1.11&  3.45& -1.05& $<$-0.5&I& $<$-5.62& $<$-3.64& $<$-10.08&    GY350               \\
 87 &$\rho$Oph-ISO 177   & 16 27 47.09& -24 45 35.10&  15.75& 12.81& 11.13&     5.2& -0.55&  3.49& -0.72& $<$-0.4&I& $<$-5.21& $<$-3.09& $<$-9.66&    GY352               \\
 88 &$\rho$Oph-ISO 178   & 16 27 49.78& -24 25 22.00&  12.78& 11.12& 10.16&     2.2& -0.65&  3.48& -0.77&    -2.0&S&   -4.60&   -2.25&    -8.78&    GY371               \\
 89 &$\rho$Oph-ISO 185   & 16 27 55.25& -24 28 39.60&  13.04& 11.59& 10.79&     1.7& -0.98&  3.45& -0.97& $<$-0.2&I& $<$-5.90& $<$-4.03& $<$-10.48&    GY397               \\
 90 &$\rho$Oph-ISO 187   & 16 27 55.58& -24 26 17.90&  10.14&  9.33&  8.90&     0.2& -0.39&  3.51& -0.62&    -5.6&S&   -3.91&   -1.32&    -7.95&    SR10/GY400          \\
 91 &$\rho$Oph-ISO 190   & 16 28  3.56& -24 34 38.60&   0.00& 15.24& 13.20&     6.1& -1.37&  3.44& -1.21& $<$-5.0&I& --& --& --&    GY450               \\
 92 &$\rho$Oph-ISO 193   & 16 28 12.72& -24 11 35.60&  13.61& 12.02& 11.09&     1.9& -1.11&  3.45& -1.05&    -1.8& I&  -5.07&   -2.89&    -9.32&    B162812-241138      \\
 93 &$\rho$Oph-ISO 194   & 16 28 13.79& -24 32 49.40&  12.35& 10.89& 10.10&     1.8& -0.64&  3.48& -0.77& $<$-0.5&S& $<$-5.19& $<$-3.06& $<$-9.60&    B162813-243249      \\
 94 &$\rho$Oph-ISO 195   & 16 28 16.73& -24  5 14.30&  10.98&  9.57&  8.86&     1.8& -0.04&  3.56& -0.41&    -0.8& S&  -4.44&   -2.03&    -8.79&    ISO 1628168-240519   \\
 95 &$\rho$Oph-ISO 196   & 16 28 16.51& -24 36 58.00&  11.31& 10.08&  9.32&     0.9& -0.55&  3.49& -0.72&    -3.2&S&   -4.31&   -1.86&    -8.43&    WSB60/B162816-243657\\

\vspace{0.3cm}

 96 &$\rho$Oph-ISO 199   & 16 28 45.60& -24 28 19.00&   9.21&  8.41&  8.00&     0.2&  0.02&  3.57& -0.38&    -1.7& S&  -4.06&   -1.53&    -8.31&    SR13                \\

\hline\hline

\# &Object&\multicolumn{2}{c} {Coordinates}& J& H& K& A$_{\rm J}$  &  Lg L$_\ast$& Lg T$_{eff}$& Lg M$_\ast$& EW (Br$_\gamma$)& Inst.&L(Br$_\gamma$)& L$_{acc}$& $\dot M_{acc}$& Other \\
&(ISO\#)& \multicolumn{2}{c}{(J2000.0)}& \multicolumn{3}{c}{(mag)} & (mag)& (L$_\odot$)& (K)& (M$_\odot$)& (\AA)& &(L$_\odot$)& (L$_\odot$)& (M$_\odot$/y)&  Names \\

\hline

97 &$\rho$Oph-ISO 059   & 16 26 31.04& -24 31  5.20&  14.96& 12.32& 10.86&     4.6& -0.49&  3.50& -0.68& $<$-0.6& S&$<$-4.06& $<$-1.54& $<$-8.13&    WL7/GY98            \\
98 &$\rho$Oph-ISO 070$^+$  & 16 27 41.61& -24 46 44.70&  17.20& 15.33& 13.55&     1.0& -3.07&  3.36& -2.23& $<$-1.2&S& $<$-4.06& $<$-3.06& $<$-9.12&    WL2/GY128           \\
99 &$\rho$Oph-ISO 075   & 16 26 51.97& -24 30 39.50&   0.00& 16.52& 13.46&    10.4& -0.69&  3.47& -0.80& $<$-3.9&S& $<$-4.06& $<$-0.40& $<$-6.92&    GY144               \\
100 &$\rho$Oph-ISO 076   & 16 26 53.47& -24 32 36.20&   0.00& 16.24& 13.12&    10.7& -0.48&  3.50& -0.67& $<$-3.0&S& $<$-4.06& $<$-0.32& $<$-6.92&    GY146               \\
101 &$\rho$Oph-ISO 085   & 16 26 58.28& -24 37 41.00&   0.00&  0.00& 14.41&    -1.0& -0.48&  3.50& -0.67& $<$-8.0&S& --& --& --&    CRBR51              \\
102 &$\rho$Oph-ISO 093   & 16 26 31.04& -24 31  5.20&   0.00& 15.65& 12.56&    10.6& -0.24&  3.53& -0.53& $<$-0.7&S& $<$-4.06& $<$-0.73& $<$-7.41&    GY188               \\
103 &$\rho$Oph-ISO 108   & 16 27  9.43& -24 37 18.80&  16.79& 11.05&  7.14&    10.5&  1.44&  3.76&  0.48&    -1.3& S&  -4.06&    1.46&    -5.84&    EL29/GY214          \\
104 &$\rho$Oph-ISO 139   & 16 27 26.29& -24 42 46.10&   0.00& 15.18& 12.66&     8.2& -0.73&  3.46& -0.82& $<$-2.8&S& $<$-4.06& $<$-0.77& $<$-7.27&    GY260               \\
105 &$\rho$Oph-ISO 161   & 16 27 37.25& -24 42 38.00&   0.00& 14.52& 11.46&    10.5&  0.27&  3.60& -0.22& $<$-0.8& S&$<$-4.06& $<$-0.30& $<$-7.17&    GY301               \\

\hline\hline\hline
\label{table_classII}

\end{longtable}

\end{landscape}

\newpage
\begin{landscape}
\begin{longtable}{clccccccccccccccl}
\caption{ Class III Objects}\\
\hline\hline
%\vskip 0.1cm
\hline\hline
(1)   & (2)   & (3)& (4)  & (5) & (6) & (7)& (8) & (9)& (10)& (11)& (12)& (13)& (14)& (15)& (16)& (17)\\ 
\# &Object&\multicolumn{2}{c} {Coordinates}& J& H& K& A$_{\rm J}$  &  Lg L$_\ast$& Lg T$_{eff}$& Lg M$_\ast$& EW (Pa$_\beta$)& Inst.&L(Pa$_\beta$)& L$_{acc}$&$\dot M_{acc}$& Other \\
&(ISO\#)& \multicolumn{2}{c}{(J2000.0)}& \multicolumn{3}{c}{(mag)} & (mag)& (L$_\odot$)& (K)& (M$_\odot$)& (\AA)& &(L$_\odot$)& (L$_\odot$)& (M$_\odot$/y)&  Names \\
\hline
\endfirsthead
\caption{continued.}\\
\hline\hline
\endhead
\endfoot

  1 &$\rho$Oph-ISO 005   & 16 25 50.53& -24 39 14.50&   9.98&  8.82&  8.33&     1.4&  0.21&  3.59& -0.26& $<$-0.5&S& $<$ -4.42& $<$ -2.01& $<$-8.86&    IRS10               \\
  2 &$\rho$Oph-ISO 011   & 16 26  3.29& -24 17 46.50&  10.67&  9.58&  9.12&     1.2& -0.18&  3.54& -0.49& $<$-0.8& I&$<$ -4.57& $<$ -2.22& $<$-8.92&    VSSG19              \\
  3 &$\rho$Oph-ISO 014   & 16 26  7.64& -24 27 41.40&  14.68& 11.85& 10.41&     5.4& -0.03&  3.56& -0.40& $<$-0.5&S& $<$ -4.64& $<$ -2.31& $<$-9.07&    B162607-242742      \\
%  4 &$\rho$Oph-ISO 016   & 16 26  9.31& -24 34 12.10&   7.74&  6.95&  6.50&     0.1&  0.61&  3.65& -0.02&     +5.7&S&    --&  --&   --&    SR3                 \\
  4 &$\rho$Oph-ISO 016   & 16 26  9.31& -24 34 12.10&   7.74&  6.95&  6.50&     --&  --&  --& --&     +5.7&S&    --&  --&   --&    SR3                 \\
  5 &$\rho$Oph-ISO 018   & 16 26 15.81& -24 19 22.10&  14.03& 11.40& 10.03&     4.8& -0.02&  3.56& -0.40& $<$-0.5&S& $<$ -4.63& $<$ -2.29& $<$-9.06&    SKS1-7              \\
  6 &$\rho$Oph-ISO 028   & 16 26 21.02& -24 15 41.50&  12.78& 10.38&  9.27&     4.5&  0.41&  3.62& -0.14& $<$-0.6&S& $<$ -4.16& $<$ -1.66& $<$-8.58&    B162621-241544      \\
  7 &$\rho$Oph-ISO 044   & 16 26 28.48& -24 15 41.20&  15.31& 12.37& 10.78&     5.4& -0.29&  3.52& -0.56& $<$-0.7&S& $<$ -4.73& $<$ -2.43& $<$-9.09&    B162628-241543      \\
  8 &$\rho$Oph-ISO 064   & 16 26 43.76& -24 16 33.30&  12.98& 10.76&  9.60&     3.8&  0.00&  3.56& -0.38& $<$-0.7&S& $<$ -4.46& $<$ -2.07& $<$-8.84&    VSSG11              \\
  9 &$\rho$Oph-ISO 066   & 16 26 44.30& -24 43 18.00&  10.99& 10.02&  9.57&     0.7& -0.51&  3.49& -0.69& $<$-0.8& S&$<$ -4.87& $<$ -2.62& $<$-9.20&    GY112               \\
 10 &$\rho$Oph-ISO 069   & 16 26 47.05& -24 44 29.90&  12.33& 11.12& 10.56&     1.3& -0.83&  3.46& -0.89& $<$-0.4&S& $<$ -5.47& $<$ -3.44& $<$-9.92&    GY122               \\
 11 &$\rho$Oph-ISO 073   & 16 26 49.23& -24 20  2.90&  12.20&  9.85&  8.69&     4.2&  0.53&  3.63& -0.07& $<$-0.4& S&$<$ -4.23& $<$ -1.75& $<$-8.72&    VSSG3               \\
% 12 &$\rho$Oph-ISO 074   & 16 26 51.12& -24 20 50.50&  13.81& 11.48& 10.21&     3.9& -0.31&  3.52& -0.57&     +1.5&S& --&  --&   --&    IRS20/GY143               \\
 12 &$\rho$Oph-ISO 074   & 16 26 51.12& -24 20 50.50&  13.81& 11.48& 10.21&    -- & --&  --& --&     +1.5&S& --&  --&   --&    IRS20/GY143               \\
 13 &$\rho$Oph-ISO 082   & 16 26 56.92& -24 28 37.10&  17.46& 14.75& 12.81&     3.8& -1.94&  3.41& -1.55& $<$-2.0&I& $<$ -5.77& $<$ -3.85& $<$ -10.12&    GY163               \\
 14 &$\rho$Oph-ISO 091   & 16 27  1.62& -24 21 37.00&  14.25& 11.07&  9.39&     6.1&  0.49&  3.63& -0.09&     +0.9&S&   --&  --&  --&    VSSG8/GY181               \\
% 15 &$\rho$Oph-ISO 113   & 16 27 11.68& -24 23 42.00&  14.21& 11.62& 10.11&     4.3& -0.31&  3.52& -0.57&     +4.0&S&   --& --&  --&    IRS32/GY228               \\
 15 &$\rho$Oph-ISO 113   & 16 27 11.68& -24 23 42.00&  14.21& 11.62& 10.11&     --& --&  --& --&     +4.0&S&   --& --&  --&    IRS32/GY228               \\
 16 &$\rho$Oph-ISO 114   & 16 27 11.71& -24 38 32.10&   0.00& 15.06& 11.06&    16.2&  1.71&  3.79&  0.64& $<$-8.0&I& --& --& --&    WL19/GY227                \\
% 17 &$\rho$Oph-ISO 135   & 16 27 22.91& -24 17 57.40&  13.33& 10.76&  9.45&     4.7&  0.26&  3.60& -0.23&     +1.3&S&  --&  --& --&    WSSG22              \\
 17 &$\rho$Oph-ISO 135   & 16 27 22.91& -24 17 57.40&  13.33& 10.76&  9.45&     --&  --&  --& --&     +1.3&S&  --&  --& --&    WSSG22              \\
 18 &$\rho$Oph-ISO 148   & 16 27 31.06& -24 34  3.20&  13.43& 11.36& 10.39&     3.6& -0.27&  3.53& -0.55& $<$-0.5&S& $<$ -4.86& $<$ -2.61& $<$-9.28&    GY283               \\
 19 &$\rho$Oph-ISO 152$^+$  & 16 27 32.68& -24 33 23.90&  16.15& 12.74& 10.90&     6.6& -0.10&  3.55& -0.45& $<$-2.0&S& $<$ -4.10& $<$ -1.58& $<$-8.31&    GY289               \\
 20 &$\rho$Oph-ISO 156   & 16 27 35.26& -24 38 33.40&  11.28& 10.23&  9.67&     0.8& -0.62&  3.48& -0.76& $<$-0.4&S& $<$ -5.27& $<$ -3.17& $<$-9.71&    GY295               \\
 21 &$\rho$Oph-ISO 158   & 16 27 36.52& -24 28 33.30&  11.99& 11.38& 11.15&     0.0& -1.29&  3.44& -1.16& $<$-1.0&S& $<$ -5.48& $<$ -3.46& $<$-9.86&    GY297               \\
 22 &$\rho$Oph-ISO 169a  & 16 27 41.49& -24 35 37.70&  14.30& 11.83& 10.56&     4.4& -0.30&  3.52& -0.57&     +0.7& S& --& --& --&    GY322               \\
 23 &$\rho$Oph-ISO 169b  & 16 27 41.64& -24 35 41.10&  14.74& 12.48& 11.26&     3.8& -0.78 &  3.46& -0.85& $<$-0.9&S& $<$-5.06& $<$-2.89& $<$-9.38&    GY322    \\
% 24 &$\rho$Oph-ISO 180   & 16 27 49.87& -24 25 40.20&   9.44&  8.12&  7.30&     1.2&  0.39&  3.62& -0.15&     +6.3& S& --& --& --&    VSSG14/GY372              \\
 24 &$\rho$Oph-ISO 180   & 16 27 49.87& -24 25 40.20&   9.44&  8.12&  7.30&     --&  --&  --& --&     +6.3& S& --& --& --&    VSSG14/GY372              \\
% 25 &$\rho$Oph-ISO 181   & 16 27 50.51& -24 39  3.10&  14.47& 12.74& 11.85&     2.5& -1.23&  3.44& -1.12&     +5.5&S& --&  --&  --&    GY373               \\
 25 &$\rho$Oph-ISO 181   & 16 27 50.51& -24 39  3.10&  14.47& 12.74& 11.85&     --& --&  --& --&     +5.5&S& --&  --&  --&    GY373               \\
 26 &$\rho$Oph-ISO 183   & 16 27 51.92& -24 46 29.60&  14.05& 11.61& 10.37&     4.3& -0.22&  3.53& -0.52& $<$-0.7& S&$<$ -4.66& $<$ -2.34& $<$-9.03&    GY377               \\
 27 &$\rho$Oph-ISO 186   & 16 27 55.65& -24 44 50.90&  12.34& 11.15& 10.47&     1.0& -0.98&  3.45& -0.97& $<$-1.5& S&$<$ -5.03& $<$ -2.84& $<$-9.29&    GY398               \\
 28 &$\rho$Oph-ISO 189   & 16 27 57.87& -24 36  2.20&  15.37& 12.98& 11.83&     4.4& -0.79&  3.46& -0.86& $<$-1.5& S&$<$ -4.86& $<$ -2.61& $<$-9.10&    GY412               \\
 29 &$\rho$Oph-ISO 192   & 16 28  5.78& -24 33 55.00&  16.75& 14.00& 12.56&     5.1& -1.08&  3.45& -1.04& $<$-2.0& I&$<$ -4.99& $<$ -2.79& $<$-9.23&    GY472               \\
 30 &$\rho$Oph-ISO 197   & 16 28 21.71& -24 42 47.10&  16.83& 14.03& 12.44&     4.9& -1.18&  3.45& -1.10& $<$-1.0& I&$<$ -5.39& $<$ -3.32& $<$-9.74&    B162821-244246      \\

\vspace{0.3cm}

 31 &$\rho$Oph-ISO 198   & 16 28 32.66& -24 22 44.90&   8.73&  7.48&  6.85&     1.3&  0.76&  3.67&  0.07& $<$-0.4& S&$<$ -4.02& $<$ -1.47& $<$-8.52&    SR20                \\

\hline\hline

\# &Object&\multicolumn{2}{c} {Coordinates}& J& H& K& A$_{\rm J}$  &  Lg L$_\ast$& Lg T$_{eff}$& Lg M$_\ast$& EW (Br$_\gamma$)& Inst.&L(Br$_\gamma$)& L$_{acc}$&$\dot M_{acc}$& Other \\
&(ISO\#)& \multicolumn{2}{c}{(J2000.0)}& \multicolumn{3}{c}{(mag)} & (mag)& (L$_\odot$)& (K)& (M$_\odot$)& (\AA)& &(L$_\odot$)& (L$_\odot$)& (M$_\odot$/y)&  Names \\

\hline

 32 &$\rho$Oph-ISO 047   & 16 26 31.04& -24 31  5.20&  14.96& 12.32& 10.86&     4.6& -0.49&  3.50& -0.68& $<$-1.0&S& $<$ -4.71& $<$ -1.34& $<$-7.93&  IRS14/GY54   \\
% 33 &$\rho$Oph-ISO 113   & 16 27 11.68& -24 23 42.00&  14.21& 11.62& 10.11&     4.3& -0.31&  3.52& -0.57&     +5.6&S&   --&  --&  --&    IRS32/GY228               \\
 33 &$\rho$Oph-ISO 113   & 16 27 11.68& -24 23 42.00&  14.21& 11.62& 10.11&     --& --&  --& --&     +5.6&S&   --&  --&  --&    IRS32/GY228               \\
 34 &$\rho$Oph-ISO 157   & 16 27 35.67& -24 45 32.62&  12.71& 11.47& 10.88&     1.4& -0.95&  3.46& -0.96& $<$-1.0& S&$<$ -5.54& $<$ -2.08& $<$-8.54&    GY296               \\
 35 &$\rho$Oph-ISO 179   & 16 27 49.97& -24 44 17.00&  13.85& 11.97& 10.94&     2.8& -0.84&  3.46& -0.89& $<$-0.7& S&$<$ -5.37& $<$ -1.94& $<$-8.42&    GY370               \\

%\hline\hline\hline
\label{table_classIII}
\end{longtable}
\end{landscape}

}

\noindent
Caption of Tables \ref{table_classII} and \ref{table_classIII}.
\newline\noindent
Column 1: running number;
\newline\noindent
Column 2: ISOCAM number from BKA01; 
a +  sign marks objects with a companion not resolved in the
2MASS photometry;
\newline\noindent
Column 3 and 4: J2000 coordinates
\newline\noindent
Column 5,6,7: 2MASS J, H, K photometry; 
\newline\noindent
Column 8: J--band extinction;
\newline\noindent
Column 9,10,11: stellar luminosity, effective temperature and mass, determined as described in the text;
\newline\noindent
Column 12: line equivalent width: negative values for emission lines;
\newline\noindent
Column 13: instrument used in the observations: S=SOFI/NTT, I=ISAAC/UT1;
\newline\noindent
Column 14: line luminosity;
\newline\noindent
Column 15,16: accretion luminosity and mass accretion rate;
\newline\noindent
Column 17: other names.

\newpage

\begin{table*}
\caption{Companions not resolved by 2MASS}
\label{table_companions}
\centering
\begin{tabular}{c l c c c c c l}     % 7 columns
\hline\hline
(1)   & (2)   & (3)& (4)  & (5) & (6)& (7) &(8)\\
\# &Object&\multicolumn{2}{c} {Coordinates}& Separation & EW& Line&  Other \\
&(ISO\#)& \multicolumn{2}{c}{(J2000.0)}&  (arcsec)&  (\AA)& &  Names \\

\hline

  1 &$\rho$Oph-ISO 002b  & 16 25 38.12& -24 22 36.30& 1.8 & $<$-1.0& Pa$_\beta$& B162538-242238  \\
 2 &$\rho$Oph-ISO 068b  & 16 26 46.43& -24 12  0.10& 3.6 & $<$-0.3& Pa$_\beta$& VSS27           \\
 3 &$\rho$Oph-ISO 070b  & 16 27 41.61& -24 46 44.70&4  &  +2.5& Br$_\gamma$&   WL2/GY128           \\
 4 &$\rho$Oph-ISO 072b  & 16 26 48.98& -24 38 25.20&  3.6 &  $<$-1.0& Pa$_\beta$& WL18/GY129          \\
% 5 &$\rho$Oph-ISO 088b  & 16 26 58.44& -24 45 31.90&  &   &    SR24S/GY167         \\
% 6 &$\rho$Oph-ISO 121b  & 16 27 15.70& -24 38 43.40&  & $<$-0.5&  WL20/GY240B         \\
 5 &$\rho$Oph-ISO 152b  & 16 27 32.68& -24 33 23.90& 3  &$<$-4.0 & Pa$_\beta$&  GY289          \\
% 6 &$\rho$Oph-ISO 169b  & 16 27 41.64& -24 35 41.10& 4 & $<$-0.9&    GY322         \\

\hline
\end{tabular}
\end{table*}

\end {document}